\documentclass[12pt]{article}
\def\slash#1{\setbox0=\hbox{$#1$}#1\hskip-\wd0\hbox to\wd0{\hss\sl/\/\hss}}
\usepackage{epsf, cite}
\usepackage{epsfig}
\usepackage[all]{xy}
\usepackage{amsfonts}
\usepackage{latexsym}

\usepackage{epsf, cite}

\makeatletter
\renewcommand\section{\@startsection {section}{1}{\z@}%
                                   {-3.5ex \@plus -1ex \@minus -.2ex}%nn
                                   {2.3ex \@plus.2ex}%
                                   {\normalfont\large\bfseries}}
\renewcommand\subsection{\@startsection{subsection}{2}{\z@}%
                                     {-3.25ex\@plus -1ex \@minus -.2ex}%
                                     {1.5ex \@plus .2ex}%
                                     {\normalfont\bfseries}}
\makeatother

\let\non\nonumber

\def\lbldef#1#2{\expandafter\gdef\csname #1\endcsname {#2}}

\def\href#1#2{#2}
\def\beq{\begin{equation}}
\def\eeq{\end{equation}}
\def    \bea    {\begin{eqnarray}}
\def    \eea    {\end{eqnarray}}
\def \pl {\partial}
\def    \be     {\begin{equation}} 
\def    \ee     {\end{equation}}
 
\def    \rt     {\right )}

\def \pa {\partial}

\def \eps {\epsilon}
\def \diag {\mbox{diag}}

\def\bea{\begin{eqnarray}}
\def\eea{\end{eqnarray}}
\newcommand{\pd}[2]{\frac{\partial{#1}}{\partial{#2}}}

\def\gs{\,\raise.15ex\hbox{/}\mkern-11.5mu G} %this one can be subscripted
\def\cs{\,\raise.15ex\hbox{/}\mkern-11.5mu C} %this one can be subscripted

\def\lt{\lfloor}
\def\rt{\rfloor}
\newcommand{\tb}[3]{[X^{#1},X^{#2},X^{#3}]}
\newcommand{\gtb}[3]{[H^*,X^{#1},X^{#2},X^{#3}]}
\newcommand{\fbr}[5]{\lt X^{#1},X^{#2},\tb{#3}{#4}{#5}\rt}

\def\p{\partial}

\def\Prp{ { {\cal{P}}_{\cal{R^+}} }  }

\def\cRp{ { { \cal{R}}^+}  }
\def\cRm{ { {\cal{R}}^-}  }

\def\fs4{ fuzzy $S^{4}$}
\def\fs3{ fuzzy $S^{3}$}
\def\fs2{ fuzzy $S^{2}$}
\def\mnc{$Mat_{N}(\mathbb{C})$} %This goes before punctuation
\def\mncs{$Mat_{N}(\mathbb{C})$ }%This has a space
\def\anstp{${\cal A}_n(S^3)$}

\providecommand{\openone}{\leavevmode\hbox{\small1\kern-3.8pt\normalsize1}}

\def\m{\begin{matrix}}
\def\em{\end{matrix}}

\def\P{\mathcal P}

\def\P{\mathcal P}

\begin{document}
\pagestyle{plain}
\begin{titlepage}

\begin{center}

\hfill{QMUL-PH-2007-01} \\

\vskip 1cm

{{\Large \bf M-theory branes and their interactions}} \\

\vskip 1.25cm {David S. Berman\footnote{email: D.S.Berman@qmul.ac.uk}}
\\
{\vskip 0.2cm
Queen Mary College, University of London,\\
Department of Physics,\\
Mile End Road,\\
London, E1 4NS, England}\\

\end{center}
\vskip 1 cm

\begin{abstract}
\baselineskip=18pt

 In recent years there has been some progress in understanding how one
 might model the interactions of branes in M-theory despite not having a
 fundamental perturbative description. The goal of this review is to
 describe different approaches to M-theory branes and their
 interactions. This includes: a review of M-theory branes themselves and their
 properties; brane interactions; the self-dual string and its
 properties; the role of anomalies in learning about brane systems; 
 the recent work of Basu and Harvey with subsequent developments;
 and how these complementary approaches might fit together.

\end{abstract}

\end{titlepage}

\pagestyle{plain}

\baselineskip=19pt

\title{Contents}

{\bf{1. Introduction}}
\bigskip

{\bf{2. M-theory branes}}

\begin{description}

\item[2.1]
Supergravity and its brane solutions

\item[2.2]
Worldvolume descriptions

\end{description}

{\bf{3. Coincident brane degrees of freedom}}

\begin{description}

\item[3.1]
Absorption Cross Section

\item[3.2]
Brane Thermodynamics and AdS/CFT

\item[3.3]
The five-brane anomaly and its cancellation

\item[3.4]
Anomalies in the Coulomb branch

\end{description}

{\bf{4. Brane Interactions}}

\begin{description}

\item[4.1]
How branes may end on branes

\item[4.2]
World volume solitions

\end{description}

{\bf{5. The self-dual string and its properties}}

\begin{description}

\item[5.1]
Absorption cross section

\item[5.2]
Self-dual string anomalies

\item[5.3]
AdS limits of the self-dual string 

\item[5.4]
Self-dual string effective actions

\end{description}

{\bf{6. The membrane boundary theory}}

\begin{description}

\item[6.1]
The boundary of the membrane

\item[6.2]
From strings to ribbbons

\item[6.3]
An effective Schild action for string ribbons

\item[6.4]
Noncommutative string field theory

\end{description}

{\bf{7. Five-Branes from membranes}}

\begin{description}

\item[7.1]
Expected properties for the membrane fuzzy funnel

\item[7.2] 
The Basu-Harvey equation

\item[7.3]
The membrane fuzzy funnel solution

\end{description}

{\bf{8. An action for multiple membranes}}

\begin{description}

\item[8.1]
Fluctuation on the funnel

\end{description}

{\bf{9. Calibrations and a generalised Basu-Harvey equation}}

\begin{description}

\item[9.1]
A linear action for coincident membranes

\item[9.2]
A generalised Basu-Harvey equation

\item[9.3]
The equation of motion

\item[9.4]
Supersymmetry

\end{description}

{\bf{10. Five-Brane configurations}}

\begin{description}

\item[10.1]
Calibrated Five-Branes

\item[10.2]
Solutions

\end{description}

{\bf{11. Fuzzy Spheres, membranes and $Q^{ 3 \over 2}$}}
\bigskip

{\bf{12. A nonassociative membrane theory?}}

\begin{description}

\item[12.1]
The Nonassociative algebra

\end{description}

{\bf{13. Other Ideas}}

\newpage

\section{Introduction}

In 1995 it was realised that the strong coupling limit of the IIA string is
an eleven dimensional theory whose low energy limit is eleven
dimensional supergravity \cite{witten1,hull1}. This unknown mysterious
eleven dimensional theory became known as M-theory. Considering that
it was a string theory at strong coupling, the properties of
M-theory were somewhat surprising. The critical dimension was eleven
not ten.  The extended objects were no longer strings but membranes
and five-branes. All the different string theories were different
compactification limits of this single theory, as such M-theory unified
string theories. The five different versions of string theory were
just M-theory expanded around different vacua. This M-theory web then
explained the nonperturbative dualities that had been conjectured in
string theory some years previously. Most elegantly perhaps, the
IIB SL(2,Z) strong weak duality was a simple consequence of M-theory
diffeomorphism invariance \cite{schwarz1}. Work took place in understanding how the
branes in M-theory fitted in with the branes of string theory and a
dictionary was uncovered between string theory objects and their
M-theory counterparts. This was very successful 
and yet there were fundamental holes in our understanding of M-theory. 
Of course, M-theory was not described by some fundamental description; 
only a definition of the low energy limit was really known. A true
formulation of M-theory away from the low energy limit is still a far 
away dream that will not be discussed here.

Yet even in the low energy limit, strong evidence emerged that there
was more to M-theory than eleven dimensional supergravity. This
evidence came from examining the branes in M-theory. 

String theory was revolutionised by the discovery of D-branes \cite{polchinskia,polchinskib}. The
understanding of these nonperturbative objects allowed a deep
connection to be uncovered between non-Abelian gauge theories and
string theory. This would result in the Maldacena correspondence
\cite{Maldacena} where
a fascinating duality between gauge theories and string/gravitational
theories has resulted in untold theoretical riches. The origin of the
non-Abelian degrees of freedom came from the open strings extending between
different D-branes becoming massless when the D-branes coincided. This
was a fundamentally stringy effect that was not apparent (at least not
immediately) from analysing the D-branes as supergravity
solutions. In fact these massless non-Abelian degrees of freedom were
apparent in the solutions in the following indirect ways. The
thermodynamic properties of the brane solutions were analysed a la Hawking and
their entropies calculated \cite{Klebanov1}. N coincident D-branes were shown to have an entropy
that scaled as $N^2$, consistent with the number of non-Abelian, U(N),
gauge degrees of freedom. The low energy scattering cross section was
calculated and again an $N^2$ scaling was found \cite{Klebanov2}. Finally anomaly
cancellation arguments also revealed the $N^2$ scaling of degrees of
freedom. All the above could be calculated without recourse to the
underlying string theory and yet the degrees of freedom themselves
were of stringy origin.

These techniques were available to analyse the branes in M-theory. The
supergravity solutions were known so one followed the D-brane example
and calculated the thermodynamics or cross section scattering or in the case of the
five-brane, its R-symmetry anomaly. All these results pointed to the
same answer\footnote{Whether these are really independent checks is
    not clear, there is a discussion of the interrelatedness of these
    results in \cite{gary1} }. The theory describing N coincident membranes has an $N^{3 \over 2}$ scaling of its
world volume degrees of freedom and the theory descibing N coincident five-branes an $N^3$
scaling. This teased the question, what are these degrees of freedom?

For the five-brane there was the guess that the origin must be in open
membranes. The membrane becomes a string after compactification
and the origin of the non-Abelian degrees of freedom on D-branes were
open strings thus these open strings were actually open membranes when
lifted to M-theory \cite{strominger1,townsend0}.

With this motivation amongst others there was a study of how
membranes may end on five-branes. This has lead to a description of open membranes as soliton like objects on the five-brane world
volume. This soliton is known as the self-dual string\cite{howe}. Its properties
have been analysed using similar techniques as for the supergravity
brane solutions. These include low energy scattering and anomaly
calculations and even the development of a Maldacena style limit on the
brane. 

Despite this, the real origin of the $N^3$ degrees of freedom of the five-brane is
still to be understood as is the $N^{3 \over 2}$ scaling of the
membrane. Yet the membrane naively seems to give greater hope. The
number of degrees of freedom is less than a gauge theory so one
might speculate that it is possible to view the membrane theory as some
sort of matrix valued field theory with constraints that restrict the real
degrees of freedom. (This is similar to many matrix models of
condensed matter systems for example.) Some eleven years after the
inception of M-theory a recent attempt was made to
construct such a model of non-Abelian membranes.
\cite{lambert1}. This was inspired by the seminal work of Basu and Harvey
\cite{harvey1} and subsequent developments \cite{berman1,
  berman2}. This theory of non-Abelian membranes is truely novel. In
order to be supersymmetric the theory has a nonassociative product for
the fields and in order for the supersymmetry algebra to close there
is a new {\it{gauge}} symmetry for the membrane fields. The evidence
for this work is still speculative but it does have the one key
property of having $N^{3 \over 2}$ degrees of freedom\footnote{A
  similar observation concerning nonassociative models and the
  membrane was made in \cite{berman2}}.

This review will follow the above narrative. We begin with
the membrane and five-brane as supergravity solutions and describe how analysing
those solutions gives us the mysterious $N^{3 \over 2}$ and $N^3$
scaling of the degrees of freedom. We will then move to world
volume descriptions and the self-dual string as a description of
membranes ending on a five-brane. We shall discuss its properties. Then
we move to describe the recent membrane models and in passing
discuss new ideas
inspired by this search for a description of interacting M-theory
branes. Unfortunately, this story does not have a final ending. Despite
eleven years of progress and the many new ideas discussed here, the
true description of M-theory branes remains unknown.

The informed reader may question the choice of topics chosen by the
author in this review since many aspects and applications of M-theory
are omitted. These include amongst others: M(atrix) theory \cite{matrixmodels}; Sieberg Witten
theory from the five-brane \cite{wqcd,l-w}; $G_2$ compactifications \cite{g2}
and; of course the basics of M-theory in unifying the string theories.

This choice came about by trying to concentrate on
aspects associated with the questions of interacting branes though
inevitably the areas in which the author is more expert receive more
attention. The final chapter on other ideas reviews areas not
covered in detail in the rest of text and hopefully there are sufficient
references throughout the review on questions not covered by the text.

\section{M-theory Branes}

\subsection{Supergravity and brane solutions}

We begin with branes as ${1 \over 2}$ BPS solutions to eleven dimensional supergravity.
The Bosonic sector of eleven dimensional supergravity (with the
coupling set to one) is described by
the following action \cite{igora,igorb}:
\beq
S = \int d^{11}x \sqrt{-g}  R - {1 \over 2} F \wedge {}^* F - {1 \over {6}}
C \wedge F \wedge F  \, \, , \label{sugra}
\eeq
where $F$ is the four form field strength of a three
form potential $C$.
Note the presence of the Chern-Simons like term. It is this term in
the action that allows the possibility of
membranes having boundaries that end on five-branes. The full
consequences of this term and membrane boundaries will be discussed in
the following section.

The supersymmetry transformation of the
gravitini (in a purely Bosonic background) is given by:
\beq
\delta \psi_\mu = (\pl_\mu  + {1 \over 4} \omega_{\mu ab} \Gamma^{ab} +
{1 \over {288}} ( \Gamma^{\alpha \beta \gamma \delta}{}_\mu - 8
\Gamma^{\beta \gamma \delta} \delta^\alpha_\mu ) F_{\alpha \beta \gamma \delta} ) \epsilon   \, \, . \label{susyvar}
\eeq
(The full action and supersymmetry variation including Fermions is described amongst
other places in \cite{igora,igorb} ).

The ${1 \over 2}$ BPS brane solutions are found by imposing a projection
on to a subspace of the spinor that generates the gravitini variation and then imposing
that the supersymmetry variation vanish. This gives a relation between the fluxes
and the spin connections. The projectors for the membrane and
five-brane respectively are:
\beq
P_{M2}= {1 \over 2}( 1 + \Gamma^{012}) \, \, ,\qquad P_{M5} = {1 \over 2} (1
+ \Gamma^{012345}) \, \, . \label{projectors}
\eeq
Solving these first order equations produces the brane
solutions\footnote{Actually one must use either the equations of
  motion for the flux or a component of the Einstein equation to fix
  the function H to be Harmonic.}. The membrane metric is given by \cite{m2metric}:
\beq
ds^2=H^{-{1 \over 3 }}(r) (-dx^0 + dx^1 + dx^2) + H^{2 \over 3}(r)
  (dr^2 + r^2 d \Omega_7^2)    \label{m2metric}
\eeq
with the flux given by
\beq
F_{01r}= \pl_r H(r)^{-1} \, \, .  \label{m2flux}
\eeq
The five-brane metric is given by \cite{guven}:
\beq
ds^2=H^{-{2 \over 3 }}(r) (-dx^0 + dx^1 + dx^2) + H^{1 \over 3}(r)
  (dr^2 + r^2 d \Omega_4^2)    \label{m5metric}
\eeq
and its flux given by
\beq
F_{\mu \nu \rho \sigma}=H(r)^{2/3} \epsilon_{\mu \nu \rho \sigma \tau} \pl^{\tau} H(r) \, \, .  \label{m5flux}
\eeq
The $\epsilon$ tensor in the above refers to the antisymmetric tensor in the 5 dimensional space transverse
to the five-brane. 
In each case the function H(r) is a harmonic function on the tranverse
space to the brane. Thus,
\beq
H_{M2}= 1 + {{2^5 \pi^2 Q_{M2} l_p^6 }\over r^6} \equiv 1+ {{R_{M2}^6} \over r^6} \qquad {\rm{and}} \qquad H_{M5}
= 1 + { { \pi Q_{M5} l_p^3} \over r^3} \equiv 1+ {{R_{M2}^3} \over r^3} \, .  \label{harmonicfunctions}
\eeq
The {\it{charge}}, Q, in the harmonic function is the
number of branes. This confirmed by simply integrating the flux
(\ref{m2flux}) or (\ref{m5flux}) over the sphere at infinity transverse to the brane world
volume.
(The reader may wonder as to the origin of the constants of 2 and $\pi$ in the above
harmonic functions. These are fixed by demanding the charge as the
calculated by the flux through the sphere at infinity be integer as
would be required by Dirac quantisation.)
These harmonic functions carry much of the information of the brane
physics and we will return to them frequently. In particular, we will
see that the important information is contained in the relationship
between, R, the length scale in the harmonic function and Q, the brane charge.
Once we have the brane
solutions then the next step is to determine their effective world volume
description.

\subsection{World volume descriptions}

The low energy dynamics of branes may be captured by an effective
world volume action. This is an action for the Goldstone
modes that are present when, due to the presence of the brane solution, the
symmetries of the supergravity action are broken. As the symmetries
are only broken on the brane these Goldstone modes are restricted to
lie in the brane world volume. Corrections to these actions will
occur when the brane radius of curvature is of the order of the
fundamental length scale of theory. For eleven dimension
supergravity that is the Plank length. Thus, the following actions are
good approximations provided the curvature is small in Plank
units. This point of view is the most conservative, where the actions
are simply low energy approximations. One might imagine that they are
fundamental actions and will receive no corrections just as for the
action of the fundamental string. The lack of
renormalisability for an arbitrary background would seem to contradict
this view.

The membrane is simple to describe from a world volume perspective. It
is simply given by a Nambu-Goto style action (ie. induced volume) with
minimal coupling to the background $C_3$ field. This gives \cite{townsend1}:
\beq
S_{M2}= {1 \over l_p^3} \int d^3 \sigma ( \sqrt{{\rm{det}}G_{\mu \nu}}
+ \pl_{\mu} X^I \pl_{\nu} X^J \pl_{\rho} X^K \epsilon^{\mu \nu \rho}
C_{IJK} )
\eeq
where the induced metric is given by:
\beq
G_{\mu \nu} = \pl_\mu X^I \pl_\nu X^J g_{IJ} \, \, .
\eeq
Of course one can also use the Howe Tucker formulation to give:
\beq
S={1 \over l_p^3} \int d^3\sigma (\sqrt{-\gamma} ( \gamma^{\mu \nu} \pl_\mu  X^I \pl_\nu
X^J g_{IJ} - 1) + \pl_{\mu} X^I \pl_{\nu} X^J \pl_{\rho} X^K \epsilon^{\mu \nu \rho}
C_{IJK}) \, \, .
\eeq
The presence of the one in the above action essentially acts like a world
volume cosmological constant. Its presence is a simple indicator that
the membrane unlike the string is not conformal. Dimensional reduction
of the above membrane action gives the string action \cite{Duff} without dilaton
coupling (to obtain the dilaton coupling one has to consider the membrane
partition function \cite{bermanperry}).

Essentially one may view the action for the single membrane as a
Goldstone mode action for the membrane solution that breaks eight
of the eleven translation symmetries of the supergravity action. Thus
there are eight physical modes on the membrane corresponding to these
modes. This is easily seen once Monge gauge has been chosen and then
the coordinates transverse to the brane become the physical fields on
the brane world volume. Their vacuum expectation values become the
classical location of the brane. Nontrivial classical solutions to the membrane
equations of motion describe different membrane embeddings and have
been studied at length in \cite{townsend}. 

The five-brane is a more difficult object to describe from a world volume
perspective. The field content (of a single five-brane) may be
determined by again examining the Goldstone modes of the solution
\cite{Cederwall}. This gives five scalars, $\phi^I, I=1..5$ corresponding to the five
broken translations and a self-dual two form potential, $b$ with field
strength $H=db$ corresponding to
normalisable large gauge transformations of the C field in the
five-brane background. There are also the Fermionic superpartners,
symplectic Majoranna Spinors. 
Together this field content forms a
(0,2) tensor multiplet in six dimensions. Due to the self-duality
constraint on H, it is not possible to write a simple action for the
tensor multiplet even at the linearised level. This maybe achieved
however by introducing an auxiliary scalar that effectively allows
one to gauge away the anti-self-dual degrees of freedom. This is known
as the PST approach \cite{pst,pst2,pst3}. The PST action for the five-brane is
given by:
\beq
S_{PST}=  \int d^6\sigma (\sqrt{-{\rm{det}}(g_{\mu \nu} + i
  {\tilde{\cal{H}}}_{\mu \nu})} - \sqrt{-g} {\tilde{\cal{H}}}^{\mu
  \nu} H_{\mu \nu \rho} v^\rho + C_6 + H \wedge C_3)
\, \, ,
\eeq
where 
\beq
{\tilde{\cal{H}}}_{\mu \nu }= { 1\over 6} g_{\mu \alpha} g_{\nu \beta}
\epsilon^{\alpha \beta \gamma \delta \rho \sigma} {\cal{H}}_{\gamma
  \delta \rho} v_\sigma
\eeq
and v is an auxiliary closed one form constrained to have unit
norm. ${\cal{H}}=db-C_3$ is the combination of the world sheet three
form field strength, with the pull back of the background three 
form $C_3$. $g_{\mu \nu}$ is the pull back of the metric to the brane and
$C_6$ denotes the pull back of the six form potential ie. the dual to
the usual three form potential of eleven dimension supergravity.
This is similar in spirit to the actions for D-branes with a Dirac
Born-Infeld type terms followed by a Wess-Zumino type term coupling to
the background fields.

Note that since the five-brane is the magnetic dual to the membrane it
couples minimally to $C_6$ the dual six form potential as opposed to $C_3$.
This action is useful for encoding the classical dynamics of the
five-brane and determining various properties such as its relation to
other branes through dimensional reduction. Many quantum aspects
though may not be captured by this action. The quantisation of
self-dual fields is a very subtle issue of which much has been
written, see \cite{wittenm5,moore} and references
therein. We will avoid discussion of these quantum aspects of the
five-brane theory but note that analysis of its anomalies and their 
cancellation will be crucial to understand properties of the coincident five-brane theory.

The equations of motion of the five-brane were first found in
\cite{howesezgin} using a doubly supersymmetric formalism. Soon
after that the PST action was found in Green-Schwarz form and then
\cite{west1} worked out the Green-Schwarz form of the equations of
motion with no auxiliary fields. The relation between these results is
described in \cite{9703127}.

The equations may be written in a useful
compact form by introducing an effective metric
\cite{gibbons2,sundell1} defined by:
\beq
G_{\alpha \beta}={{ 1+K} \over {2 K}}( g_{\alpha \beta} + { {l_p^6} \over 4}{\cal{H}}_{\alpha \mu \nu} g^{\mu
  \sigma} g^{\nu \rho}
{\cal{H}}_{\beta \sigma \rho}  ) \, \, ,   \label{OMmetric}
\eeq
where
\beq
K=\sqrt{1+{l_p^6 \over 24} {\cal{H}}^2} \, \, .
\eeq
Using this metric one may write the equation of motion for the scalars
as \cite{gibbons2,sundell1}:
\beq
\pl_\mu \sqrt{G} G^{\mu \nu} \pl_\nu \phi^I =0 \, \,  \quad I=1..5 \, .
\eeq
The equation of motion for the three form field strength becomes a
nonlinear self-duality equation given by:
\beq
{1 \over 6} \sqrt{-{\rm{det}}g} \epsilon_{\mu \nu \rho \sigma \lambda
  \tau} {\cal{H}}^{\sigma \lambda \tau} = {1 \over 2} (1 +K)
(G^{-1})_\mu {}^{\lambda} {\cal{H}}_{\nu \rho \lambda} \, .
\eeq
In the weak field limit this is simply $H=*H$ but for general field
configurations this is a complicated nonlinear theory.

The various non-trivial solutions to these equations yield a
variety of five-brane configurations embedded in spacetime. They 
also can describe membranes ending on a five-brane. This is the
self-dual string solution \cite{howe}. It is this that will be
analysed in detail as a way of describing how
membranes may end on five-branes.

In the above we have really only dealt with Bosonic aspects of the
equations of motion. It is worth mentioning however that it is stll
not technically feasible to give the M-brane actions or equations of
motion, indeed any brane action apart from the string, in component
form in a generic background. That is in a form where the full theta 
expansion of the target space superfields are worked out. The target
spaces where it is known to be possible are flat space or AdS $\times$
a sphere. The reader may consult \cite{0011173} for when one wishes to study
more general cases.

\section{Coincident Brane Degrees of freedom}

Returning to the supergravity brane solutions one may attempt to
analyse their properties. In particular, we are interested in knowing
how many degrees of freedom will be present when we have N coincident
branes; N is synonymous with the charge Q in the brane harmonic functions 
(\ref{harmonicfunctions}). 

We must ask first of all what we mean by degrees of freedom. There
will be three notions that we will use. The first will be thermal entropy
of the system. Branes have horizons and so just like black holes they have thermal
properties. This may be determined a la Hawking \cite{hawking}. 
One then has a notion of entropy given by the usual
laws of blackhole thermodynamics. The scaling of the entropy with
N will determine how the number of degrees of freedom scale with the
number of branes. We can also use the AdS/CFT correspondence to
calculate thermodynamics of the decoupled brane theory
using the properties of black holes in AdS. This is equivalent to the
black brane picture in the decoupled limit.

Secondly, we can consider a low energy scattering calculation where
one examines the infrared fluctuations of, for example, a graviton 
in the background of a brane solution. From this one may determine 
an absorption cross section for the brane solution. This absorption
cross section will scale with N, the number of branes. The absorption
cross section implicitly also measures the number of degrees of
freedom since the more degrees of freedom an object carries
the greater the absorption probability. Thus, the scaling of the absorption
cross section with N measures how the degrees of freedom scale with
N. Of course this calculation really only measures the massless degrees of freedom since
those are the only modes available to the infrared
fluctuations. (It is the massless modes which are those of interest
anyway). Since the absorption cross section effectively gives a transverse area of
the brane it is no surprise then that this agrees with the
thermodynamic calculation since the entropy of a black hole famously
scales as its area. In fact, the universal nature
of these calculations has
been discussed in \cite{gary1} and so we might wonder whether this
is really an independent notion/calculation of the degrees of
freedom. The view taken here is that this just confirms that they
are both sensitive to the same physics.

Thirdly, for the five-brane we will have the power of anomalies at our
disposal \cite{harvey2}. This notion is more akin to what one does in string theory
where the central charge measures the number of degrees of freedom and
is determined via the Weyl anomaly. For the five-brane the procedure
is as follows. One first calculates the anomalies of the world volume
theory for a single
brane. Then one cancels the anomaly from terms in the
supergravity action; this cancellation works via the so called {\it{inflow}}
mechanism and with some care taken with the Chern-Simons term. 
One then determines how the anomaly cancellation terms in the
supergravity action scale
with N, the number of branes. This then determines how the world
volume anomaly itself scales
with N. (Of course here there is an assumption that the anomaly will
always cancel, independently of N; that is an assumption but a weak
one; since if it didn't cancel then M-theory would not be quantum mechanically
consistent.) Now, once one knows how the R-symmetry anomaly scales
with N, then supersymmetry implies the Weyl anomaly scales the same
way since they are part of the same {\it{anomaly}} multiplet. The
scaling of the Weyl anomaly gives the central charge and the effective
number of degrees of freedom. 

Unfortunately, the membrane being a
three dimensional theory is free of local anomalies and so we can't use this
technique in this case. It is still an open question, though, how
one interprets the string Weyl anomaly from the membrane point of
view. For the relationship between the Weyl symmetry of the string
and the membrane see \cite{Stelle}.

\subsection{Absorption Cross Section}

This was the first indication that coincident M-theory branes had
an unusual scaling of the number internal degrees of freedom. One
simply calculates the scattering amplitude of some field off the
brane solution in supergravity and then work out the absorption cross
section \cite{Klebanov2}. This will depend on the brane degrees of freedom.
It is sufficient to consider the S-wave of a minimally coupled scalar
in the brane background. This then reduces to a simple Coulomb like
problem. In the presence of a five-brane a minimally coupled scalar field $\phi(\rho)$ obeys:
\beq
\left(\rho^{-4} { d \over {d \rho}} \rho^4 {d \over {d \rho }} +1 +
 {{ (\omega R_{M5} )^3} \over {\rho^3}} \right) \phi(\rho) =0
\eeq
where $\rho = r \omega$ the dimensionless radius, $\omega$ is the energy and $R_{M5}$ is
defined in (\ref{harmonicfunctions}). The solution in the inner
region, $\rho << 1$ is:
\beq
\phi=iy^3(J_3(y)+iN_3(y))
\eeq
where $y={2 (\omega R)^{3/2}
\over {\sqrt{\rho}}}$. This can be matched onto the solution in the
outer region, where $\rho >> (\omega R)^3$:
\beq
\phi=24\sqrt{{2 \over \pi}} \rho^{-3/2} J_{3/2}(\rho)
\eeq
giving an absorption probability of 
$ {\cal{P}}={\pi \over 9} (\omega R)^9$. This finally yields an
absorption cross section of 
\beq
\sigma_{M5}= {2 \over 3} \pi^3 \omega^5 R_{M5}^9 \sim N_{M5}^3 \, .
\eeq
A similar calculation for the membrane yields:
\beq
\sigma_{M2}= {2 \over 3} \pi^4 \omega^2 R_{M2}^9 \sim N_{M2}^{3/2} \, .
\eeq
Note that the cross section in both cases goes like $R^{9}$; it is
only the dependence of R on the number of branes that changes between
the membrane and five-brane.

\subsection{Brane thermodynamics and AdS/CFT}

The AdS/CFT correspondence \cite{Maldacena} is an invaluable tool for
examining CFT at strong coupling. It provides us with some insight into
the decoupled CFTs appearing on M-theory branes. It is the
shortest route to examining the number of degrees of freedom carried by
a brane. Going via the AdS/CFT correspondence rather than directly looking at
the brane thermodynamics is beneficial because one can also see other
properties of the decoupled brane theory.

The starting point for the AdS/CFT correspondence is to take a low
energy limit and simultaneously go to the near horizon of the solution such that the
supergravity action remains finite in the limit.
These considerations give the following limits:
\beq
l_p \rightarrow 0 \qquad r \rightarrow 0 \qquad {r^2 \over l_p^3} =u \quad
{\rm{fixed}} \label{m2limit}
\eeq 
for the membrane and:
\beq
l_p \rightarrow 0 \qquad r \rightarrow 0 \qquad {r \over l_p^3} =u^2 \quad
{\rm{fixed}}
\eeq
for the five-brane. In these limits the membrane spacetime becomes $AdS_4 \times
S^7$ and the five-brane spacetime becomes $AdS_7 \times S^4$. The key information
is contained in the radius of the AdS spaces. From (\ref{harmonicfunctions}) one sees that 
\beq
R_{AdS_4}={1 \over 2} (2^5 \pi^2)^{1 \over 6} N^{1 \over 6} l_p \label{m2rad}
\eeq
with $2 R_{AdS}= R_{S^7}$ for the membrane and 
\beq
R_{AdS_7}=  2 \pi^{1/3} N^{1 \over 3} l_p   \label{m5rad}
\eeq
with $R_{AdS}= 2 R_{S^4}$ for the five-brane. 

We will follow \cite{wittenads} to use AdS black holes to determine
thermodynamics of the dual theory. Blackhole solutions whose asymptotics are $AdS_{n+1}$ are
given by:
\beq
ds^2= -(1 + {r^2 \over l^2} - {{M w_n} \over r^{n-2}}) dt^2+ (1 + {r^2
  \over l^2} - {{M w_n} \over r^{n-2}})^{-1} dr^2 + r^2 d \Omega^2
\eeq
where
\beq
w_n= { {16 \pi G_N } \over {(n-1) Vol(S^{n-1})}}  \, ,
\eeq
and $Vol(S^{n-1})$ refers to the volume of the unit sphere of
dimension $n-1$. $l^2$ is related to the radius of the AdS space by : $l^2= R_{AdS} l_p $. 
M has the interpretation of the mass, as defined using some ADM like
prescription with a subtraction scheme at infinity. The subtraction
scheme is needed since the space is asymptotically AdS and a naive application of the ADM prescription
would yield an inifite result.

For the cases relevant to M-theory we have n=3 and 6
corresponding to the membrane and five-brane respectively.
There is obviously a horizon whose radius is given by the largest
root of the equation:
\beq
1 + {r^2 \over l^2} - {{M w_n} \over r^{n-2}}=0 \, \, .
\eeq
This is quite complicated for general $n$ however in the large
M limit the horizon radius, $r_h$ will scale as:
\beq
r_h \sim (w_n l^2)^{1 \over n} M^{1 \over n} \, .
\eeq
The temperature is given by:
\beq
T_{bh}= { {n r_h^2+((n-2)l^2} \over {4 \pi l^2 r_h}} \, .
\eeq
We will find it useful to consider the large mass or equivalently the
large horizon limit limit where:
\beq
T_{bh} \sim {{r_h} \over {l^2}} \,. \label{temp}
\eeq
This is true independent of the dimension (the constant of
proportionality will change however).
The entropy is given by the usual Hawking area formula:
\beq
S= {{\rm{Area}} \over {4 G_N}} \, . \label{hawk}
\eeq
Now we must be careful with what we mean by $G_N$. This will be $G_N$
in the $AdS_{n+1}$ space where the black hole is. The relation to
the eleven dimensional $G^{(11)}_N$ is given by dividing by the volume
of the $11-(n+1)$ dimensional sphere upon which eleven
dimensional supergravity has been reduced. Since the radius of the
sphere is proportional to $R_{AdS}\sim l^2$ we have:
\beq
G_N \sim {{G_N^{(11)}} \over {(l)^{2(10-n)}}} \, .  \label{gnewt}
\eeq 
The area of the black hole will scale as $r_h^{n-1}$ which using
(\ref{temp}) implies the area scales as $l^{2(n-1)}
T^{n-1}$. Combining this expression with (\ref{gnewt}) gives the entropy
in the canonical ensemble as:
\beq
S \sim l^{18} T^{n-1} \sim R^9 T^{n-1}  \, .
\eeq
Note, as with the absorption cross section, the scaling with l or 
equivalently R is independent of the dimension. We now use
the relationship between the $AdS$ radius and the number of
branes (\ref{m2rad},\ref{m5rad}); this will be dimensionally dependent to give the
entropies for the membrane and five-brane respectively:
\beq
S_{M2} \sim N^{3 \over 2} T^2 \qquad S_{M5} \sim  N^{3} T^5 \, .
\eeq
These thermodynamic relations relating the temperature to entropy are
inevitable for any conformal theory since there is no scale and the
entropy is expected to be extensive; this fixes the temperature
dependence. The N dependence though is not determined by dimensional
arguments. This factor is governed by the number of degrees of
freedom present in the system. One might wander what would happen if
we moved away from the {\it{large}} black hole limit. These
corrections may be calculated and are associated to finite size effects in the
conformal field theory \cite{gubser}.

\subsection{The five-brane anomaly and its cancellation}

The (0,2) tensor multiplet world volume theory of the five-brane is
anomalous. There are two sources of anomalies. The chiral two form is anomalous under world volume
diffeomorphisms and the Fermions have an SO(5) R-symmetry 
anomaly. (In more formal terms, the SO(5) R-symmetry acts as SO(5) gauge
transformations on the normal bundle N to the brane and the
diffeomorphisms act as local SO(1,5) transformations on the tangent
bundle). The total anomaly may be determined through descent by an eight form characteristic class,
which for the world volume theory we denote as, $I^{WV}_8$. (For a relevant review of
anomalies see \cite{alvarezgaumewitten}).
That is $I^{WV}_8$ determines the anomaly $I^1_6$, via
\beq
I^{WV}_8=d I^0_7 \qquad \delta I^0_7= d I^1_6  \, ,
\eeq
where $\delta$ indicates the variation under which the action is
anomalous. The specific eight form may be read off using
\cite{alvarezgaumewitten} and is described in \cite{wittenm5}.
The anomaly is cancelled by a combination of an inflow mechanism and a
careful treatment of the Chern-Simons term. 
The inflow mechanism is as follows. There is an interaction term 
present in the supergravity action at one loop in
$l_p$.
\beq
S_{inflow}=F_4 \wedge I^{inflow}_7=C_3 \wedge I^{inflow}_8 \, ,
\eeq
with $I^{inflow}_8=dI^{inflow}_7$.
Taking the variation of this implies:
\beq
\delta S_{inflow} = F_4 \wedge \delta I^0_7 = F_4 \wedge d I^1_6 = -d F_4
\wedge I^1_6 \label{m5anom} \, .
\eeq
In the presence of a five-brane
\beq
d F_4 = Q_5 \delta^6     \label{m5charge}
\eeq
where $\delta^6$ is essentially a Dirac delta like object restricting
the form to lie on the five-brane. It is a representative of the
five-brane Thom class \cite{Bott}.
One then sees that after inserting (\ref{m5charge}) into
(\ref{m5anom}) and integrating one produces a term on the five-brane
world volume.
The sum of the inflow term and the world volume anomaly is given by:
\beq
I^{WV}_8+ I^{inflow}_8= {{p_2(N)} \over {24}}  \, ,
\eeq
where $p_2(N)$ denotes the second Pontryagin class of the normal bundle
N. The tangent bundle anomaly has vanished but the normal bundle anomaly
remains \cite{wittenm5}. The remaining contribution comes from a careful treatment of
the M-theory Chern-Simons term when there are five-branes present \cite{fhmm,fhmm2}.

This is quite subtle, the essential idea is that in the presence of
five-branes the Chern-Simons term is modified and becomes anomalous.
We introduce $\rho$, the integral of a bump form $d\rho$, and $e^0_3$
related to the global angular form ${1 \over 2} e_4$ via $d
e^0_3=e_4$. We define $\sigma_3= {1 \over 2} \rho e^0_3$. Then the
Chern-Simons term becomes \footnote{Here we use the conventions of
    \cite{fhmm,fhmm2} for the Chern-Simons term since the application of the
    Bott and Cattaneo formula is more immediate.}:
\beq
S_{CS}= {\rm{Lim}}_{\epsilon \rightarrow 0} -{3 \over \pi}
\int_{M^{11}- D_\epsilon(M^6)} (C_3 -  \sigma_3)\wedge d (C_3 -  \sigma_3) \wedge d(C_3 - \sigma_3)
\eeq
where $D_\epsilon(M^6)$ denotes the excision of a neighbourhood of
radius $\epsilon$ around the five-brane world volume $M^6$.
The variation of this term under SO(5) normal bundle gauge
transformations may be calculated using a result of Bott and Cattaneo
to yield (again expressed through descent) \cite{fhmm}:
\beq
I^{CS}_8=-{{p_2(N)} \over {24}}
\eeq
which cancels the remaining normal bundle anomaly. 

All this has been carried out for a single five-brane. The interesting
thing happens if we consider the case of $Q_5$ coincident branes and see
how these terms scale with $Q_5$.
The key point is that although the inflow term is linear in $C_3$ the
Chern-Simons term is cubic and thus will scale as $Q_5^3$. We can now 
determine the anomaly of the general five-brane world volume theory by insisting
 that the total theory is anomaly free ie. the anomaly cancellation
persists for any number of coincident branes. Since we know how the terms that cancel the
world volume anomaly scale with $Q_5$ we can infer that the world 
volume anomaly, $I^{WV}(Q_5)$ for $Q_5$ five-branes must be:
\beq
I^{WV}(Q_5)=Q_5 I^{WV}|_{Q_5=1}+{1 \over {24}} (Q_5^3-Q_5) p_2(N) \label{anomcanc}
\eeq
This is an example of the extraordinary power of anomalies. By
demanding anomaly cancellation we see that the mysterious unknown
world volume theory has a normal bundle anomaly scaling as
$Q_5(Q_5^2-1)$ To leading order in $Q_5$, this is the usual $Q_5^3$
term. The R-symmetry anomaly is in the same supermultiplet as the Weyl
anomaly and so the Weyl anomaly will scale the same way. The Weyl
anomaly provides the central charge of theory and is an
independent measure of the number of degrees of freedom. It is
satisfying that the various different methods agree to leading order
in $Q_5$. The gravity description is not valid away from that limit though
the anomaly calculation is valid for any $Q_5$.

The membrane theory is non-anomalous and so we can't use anomaly
arguments there.

The Weyl anomaly has also been calculated via the AdS/CFT
correspondence by \cite{manskostas}. This again yields the $Q_5^3$
dependence but now to get the next to leading term one must go to next
order in the $l_p$ expansion \cite{Tseytlin}.

\subsection{Anomalies in the Coulomb branch}

A more sophisticated five-brane set up may be considered where the
five-branes are in the Coulomb branch and the different five-branes
{\it{knit}} together. That is they may wind around each other with
$|\phi|=c$, fixed but $\phi^a(\sigma^\mu)$ a function of the five-brane
world volume.
Also the possibility of a more general {\it{gauge group}} 
may be considered along with different breakings of
this group{\footnote{So far we have
    not really mentioned a specific group associated to the brane
    theory but since theory under a toroidal reduction goes to
    Yang-Mills theory with gauge group G, it is natural to associate a gauge
    group to the unknown five-brane theory; the default being, as with
    D-branes, SU(N).}}.
This was studied and
the anomaly cancellation analysed by \cite{GanorMotl,Intrill}. Here we
will follow \cite{Intrill}. 

We consider the case where the mysterious interacting five-brane theory
associated to some group G is broken to a subgroup H by the vacuum
expectation values of the scalars as follows:
${\bf{\phi}}^a_i=\phi^a(T)_{ii}$ where the $(T)_{ii}$ are the diagonal
components of a generator of the Cartan of G whose little  group is
$H \times U(1)$. The massless spectrum of the world volume theory when
$<\phi^a>\neq0$ is the (0,2) theory now with group H and the tensor
multiplet associated to the U(1). The $\phi^a$s are the scalars in the
tensor multiplet. One would then expect that for energies much less than
the vev ie. $E<<(\phi^a \phi^a)^{1/4}$ these theories would decouple
and the U(1) multiplet become free.

Importantly, in order to ensure 't Hooft anomaly matching between 
theory with group G and the broken
theory with group $H \times U(1)$ there is a Wess-Zumino interaction
term on the five-brane world volume. This term is independent of the
value of the scalar vev and persists at all energy scales. Since it
is a topological term and its coefficient is quantised, it won't be
renormalised.

For $|\phi| \neq 0$ the configuration space of the scalar fields will
be  $SO(5)/SO(4) = S^4$
with coordinates given by ${\hat{\phi}}^a= {{\phi^a} \over {|\phi|}}$. The
required Wess-Zumino term is calculated to be \cite{Intrill}:
\beq
S_{WZ}={1 \over 6}(c(G)-c(H))\int_{M^7} \Omega_3({\hat{\phi}},A)
\wedge d \Omega_3({\hat{\phi}},A) \, .  \label{wzm5}
\eeq
$M^7$ is the seven dimension space whose boundary is the five-brane
world volume. $\Omega_3({\hat{\phi}},A)$ is defined via the equation
$d \Omega_3({\hat{\phi}},A) = \eta_4 = {1 \over 2} e_4^M$ where
$e_4^M$ is the pull back to $M^7$ via ${\hat{\phi}} : M^7 \rightarrow
S^4$ of the global angular, Euler class 4-form $e_4$ which enters in
the usual anomaly cancellation mechanism. ($A$ denotes the SO(5) connection). 
The coeffeicient $c(G)$ in (\ref{wzm5}) is the term in front of the normal bundle
cancellation term ie. the coefficient of the ${{p_1(N)} \over {24}}$
term in (\ref{anomcanc}). For the case considered above, corresponding to
the SU(N) theory, it is given by, $c(G)=Q_5(Q_5^2-1)$. Intriligator
conjectures that for a general group:
\beq
c(G)=|G| C_2(G)
\eeq
where $C_2(G)$ is the dual Coxeter number, normalised to N for $SU(N)$.
It would be good to have an independent check of this conjecture but
so far there are none.

There is also a Wess-Zumino term coupling to the three form field
strength, H given by \cite{Intrill}:
\beq
S_{WZ}= \alpha(Q_5)  \int_{M^7} H_3 \wedge d \Omega_3({\hat{\phi}},A)
\label{WZ2} \, ,
\eeq
with 
\beq
\alpha(Q_5)= {1 \over 4} ( |G| -|H| -1) \, . \label{alpha}
\eeq
where $|G|$ denotes the dimension of the group, G, which depends on $Q_5$
Again, this is derived somewhat indirectly. We will see later that
this term (\ref{WZ2}) is crucial for anomaly cancellation
of the self-dual string and without it the self-dual string would have
a normal bundle anomaly.

Other anomaly considerations such as for five-branes at fixed points of
orbifolds have been studied in \cite{Yi}.

\section{Brane Interactions}

So far we have explored the properties of membranes and five-branes
on their own. We now wish to examine how they may interact. The
fundamental interaction is via the membrane ending on the
five-brane. From this point of view the five-brane is the Dirichlet
brane for the membrane; a sort of D-brane for M-theory.

\subsection{How can branes end on branes}

The realisation that the membrane may end on a five-brane is due to
Townsend \cite{townsend0,Townsend3} and Strominger \cite{strominger1}. This would also be expected from
dimensional reduction arguments. Since the membrane reduces to a
fundamental string and the five-brane to a D4-brane, the fact that a
fundamental string may end on a D4 implies a membrane should be able
to end on a five-brane once theory is decompactified to eleven
dimensions.

To see directly how the membrane may end on a five-brane one examines the Chern-Simons
term in the supergravity action.
\beq
S_{CS}=\int C_3 \wedge F_4 \wedge F_4
\eeq
Even though the potential, $C_3$, appears in the action, this term 
is gauge invariant in the absence of boundaries via the
usual Chern-Simons argument.
\beq
\delta_{\lambda} S_{CS}=\int d \lambda \wedge F_4 \wedge F_4 = \int d(\lambda
\wedge F_4 \wedge F_4 )
\eeq
Thus the variation is a simple boundary term.
Now because of this term the equations of motion for $C_3$ are:
\beq
d^*F_4+F_4 \wedge F_4 =0
\eeq
which one may write as
\beq
d( ^*F_4 + F_4 \wedge C_3)=0 \, .
\eeq
Thus the charge of the membrane is actually calculated by:
\beq
Q_2=\int_{M^7} {}^*F_4 +F_4\wedge C_3   \, . \label{m2charge}
\eeq

For an infinite membrane $M^7$ would be some seven cycle enclosing the
membrane capturing all the flux. However if the membrane has a
boundary then the seven cycle maybe {\it{slipped}} off the end of the
membrane and contracted. One needs to consider the presence of a five-brane at the
boundary of the membrane. The seven cycle $M^7$ would now decompose
into a product $M^4 \times M^3$ where the $M^4$ is the four cycle
enclosing the five-brane (which we take to be infinite) and the $M^3$
is a three cycle enclosing the boundary of the membrane inside the
five-brane world volume. Let us examine the second part of this integral (\ref{m2charge}).
The integral becomes:
\beq
Q_2=\int_{M^4} F_4 \int_{M^3} C_3 = Q_5 \int_{M^3} C_3
\eeq
For now take $Q_5=1$ corresponding to a membrane ending on a single
five-brane. We see that the membrane charge must be given by 
\beq
Q_2=\int_{M^3} C_3
\eeq
where $M^3$ is a cycle surrounding the string that is the membrane
boundary inside the five-brane world volume. Thus, for a membrane to
end on a five-brane requires the five-brane to carry a nontrivial $C_3$
field. Note that since $M^3$ is closed this expression for the charge is gauge invariant as it
must be.

This is entirely in terms of the supergravity fields. As we have seen
the five-brane has a world volume two form field corresponding to the
Goldstone modes of $C_3$. The above integral in terms of the world
volume field would yield, 
\beq
Q_2= \int_{M^3} H   \label{SDstringcharge} \, ,
\eeq
where H is the field strength of the Goldstone field in the five-brane
world volume.

Powerfully, one may use the fact that the membrane may end on a
five-brane to derive the five-brane equations of motion. Just as
$\kappa$-symmetry of a closed membrane allows one to determine the
supergravity equations of motion, the requirements of
$\kappa$-symmetry for an open membrane allows one to determine 
the five-brane equations of motion \cite{chusezgin} (the
five-brane field being the backgound fields for the membrane boundary).

\subsection{World Volume solitons}

The above argument demonstrates how a world volume field of the
five-brane with nontrivial flux gives rise to membrane charge. This
argument is somewhat cohomological and does not actually yield a
membrane description. To do this we must solve the five-brane
equations of motion and find a solution with a charge
(\ref{SDstringcharge}).
We begin with the equations of motion for the five-brane. We will look
for a solution that corresponds to a membrane ending on a five-brane
and so we expect this to be a string solution from the five-brane
perspective. Thus we make an anstatz where the five-brane world volume
fields are independent of time and $x^1$ the string direction. The
remaining rotational symmetry implies that the fields may only depend
on $r$, where $r$ is the radial coordinate of the space transverse to the
string ie. $r^2= x_2^2 +x_3^2 +x_4^2 +x_5^2$.

The solution we are looking for will be somewhat like a string
monopole since it will be charged magnetically as given by
(\ref{SDstringcharge}). The simplest monopole-like solutions will be
supersymmetric ${1 \over 2}$ BPS states of the five-brane world volume
theory. Thus the simplest approach is to take the five-brane
supersymmetry transformations and search for a half supersymmetric
solution. This is just like looking for a brane solution in
supergravity, one makes an ansatz for the fields and then searches for
solutions that preserve half the supersymmetry variations.

Following the intuition gained from brane solutions in supergravity
and monopole solutions in gauge theories
one picks the supersymmetry projector to be:
\beq
\pi_{string}={1 \over 2}(1+ \gamma^7 \gamma^{01}) \label{stringproj}
\eeq
where $\gamma^i$ are five-brane world volume gamma matrices and
$\gamma^7$ is a gamma matrix in the transverse space; that is it acts
on the spin cover of the SO(5) R-symmetry of the five-brane world
volume theory. Note, that it
is essentially the same as the membrane projector for a membrane whose
world volume lies along the 017 directions. The 01 directions being common
to the five-brane and the 7 direction orthogonal.

Using this projector and the field ansatz where all field are only
functions of $r$ reveals the following from the demanding the
supersymmetry variation on the five-brane vanishes:
\beq
H= *_4 d \phi  \label{stringbps}
\eeq
where $*_4$ denotes Hodge duality in the four transverse directions to
the string . The SO(5) R-symmetry has been used to fix a single scale
to be excited.
Using the Bianchi identity for H then implies:
\beq
dH=d*d \phi(r)=0 
\eeq
which implies $\phi$ is a harmonic function. Its solution is given by:
\beq
\phi(r)= { {2 Q_{SD} l_p^2 } \over r^2 }  \equiv {R_{SD}^2 \over r^2} \, \, \label{sdstringsolution}
\eeq
with the field strength H related to $\phi$ by
(\ref{stringbps}) and H obeying the self duality relation. Note this
solution obeys, $H=*H$ even though the field equation is nonlinear. 
This is typical of a BPS solution where the nonlinearity
becomes linearised in some sense.

$Q_{SD}$ refers to the self-dual string charge and via (\ref{SDstringcharge})
also gives the numbers of membranes ending on the five-brane world
volume.

This solution will obviously be a brane in its own right and as such
there will be Goldstone modes associated to it and one might imagine
constructing an effective world volume theory describing its low energy
dynamics. 
The obvious Bosonic modes will be the four
scalars coming from the four broken translation symmetries of the
string solution in the six dimensional five-brane world volume. The
Fermionic superpartners of these Bosons will be charged under
\beq
SO(1,1)\times SU(2) \times SU(2) \times SU(2) \times SU(2) \, \, , \label{sdfermionsym}
\eeq
the first two SU(2) form an SO(4) which is
the symmetry of the space transverse to the string but parallel to the
five-brane worldvolume. The second two SU(2)s form the SO(4) that is
the symmetry of the space transverse to both the membrane and five-brane. 
From analysing the projector (\ref{stringproj}) one may determine the
supersymmetry of the Goldstone modes. There will be (4,4)
supersymmetry in two dimensions with the Fermions in the following
representations of (\ref{sdfermionsym}):
\bea
&+{1 \over 2},&1,\quad 2,\quad 2,\quad 1 \\   \nonumber
&-{1 \over 2},&2,\quad 1,\quad 1,\quad 2    \, . \label{sdfermionreps}
\eea
We thus now have a description of the field content of the self-dual string. 
In what follows we will analyse the self-dual string and its
properties just as we have analysed the properties of the membrane and
five-brane itself. 

First a word of warning, the self-dual string will couple to the
self-dual two form potential, $b_2$ in the five-brane world volume. The coupling
constant in a self-dual theory must be fixed to be of order one and so
it is never in a perturbative regime. This means one must be very
careful in determining when a classical approximation is valid and a
low energy effective description can make sense.

The self-dual string is thus a mini-version of an M-theory brane. We
know the field content from analysing Goldstone modes and can as we
will see determine many properties indirectly but we do not have a
{\it{fundamental version}} of theory. How self-dual strings
interact and how to describe coincident self-dual strings will be yet
another M-theory mystery. 

There have also been solutions found that are non-BPS by brute force
approaches to solving the nonlinear equations \cite{malcolm}. This
solution involves the two form potential only and does not excite any
of the world volume scalars. In fact
a whole family of non-BPS solutions were found using a solution generating
symmetry of the five-brane equations of motion \cite{berman3}. This
family of solutions interpolates between the solution found in
\cite{malcolm} where no scalar field was excited and the BPS solution
where the scalar field becomes singular at the origin. These are
bump like solutions which being non-BPS will undoubtedly decay to the BPS
state with the same membrane charge. This process has so far not been studied
but would be an interesting topic for future research.

\section{The self-dual string and its properties}

We will follow our intuition of how we explored the five-brane and
membrane by using techniques such as: scattering cross
sections, anomalies and various limits. First, let us state our goals. We have a
solution of the five-brane equations of motion and we would like to
know how many degrees of freedom that object carries as a function of
the five-brane charge and the self-dual string charge. This is
analogous to our original questions concerning the membrane and
five-brane given in previous sections. Since ultimately the self-dual
string describes open membranes from the five-brane perspective it is
hoped that this will throw light on our original question on the
degrees of freedom in M-theory. 

\subsection{Absorption Cross section}

The absorption cross section for s-wave scalar fluctuations is calculated 
at low energy by expanding the five-brane equations of motion around the
self-dual string background. This gives the following Coulomb type
equation for the fluctuations, $\phi(\rho)$, where we have introduced
the dimensionless radius $\rho=\omega r$ as before:
\beq
\left( \rho^{-3} {d \over {d\rho}} \rho^3 {d \over {d \rho}} + {{(R_{SD} \omega)^6}
  \over {\rho^6}}  \right) \phi(\rho)  =0  \, .  \label{flucteom}
\eeq
Solving this equation in the exterior region where $(R\omega)^6 <<
\rho^4$ gives:
\beq
\phi(\rho)= \rho^{-1} (A J_1(\rho) + B N_1(\rho))  \label{solutions}  \, .
\eeq
The interior region ($\rho << R \omega   $) has solution:
\beq
\phi(\rho)= A' cos\left( { { (R \omega)^3} \over { 2\rho^2} } \right)
+ B' sin\left( { { (R \omega)^3} \over { 2\rho^2} } \right) \, .
\eeq  In the overlap region $R \omega >> \rho >> (R \omega) ^{3/2}$
we can match the solution and so calculate the transmission coefficient.
 
This gives an absorption cross section for
the self dual string to be \cite{berman4}:
\beq
\sigma \sim R_{SD}^3 \omega^3 \sim Q_2 l_p^3 \omega^3
\eeq
This indicates that theory of $Q_2$ coincident self-dual strings
has a $Q_2$ scaling of degrees of freedom. This calculation however
does not reveal anything about the dependence of self-dual string on the
five-brane charge $Q_5$ since we could only calculate the cross section
using theory of a single five-brane. To do this we resort to an
anomaly calculation analogous to that of the five-brane itself.

This is a half BPS state. One can also consider a system with less
supersymmetry. One suggestion is the emergence self-dual string
webs, \cite{KYLee}. This is when the strings end on each other forming a
net or web like structure.
This would correspond to the situation of
intersecting membranes ending on the five-brane. The supersymmetry
restricts the possible angles allowed for membrane intersections and
equivalently, for the tension of the string web to balance, the string
vertices must intersect at particular angles.

\subsection{Self-dual string anomalies}

As can be seen from the table of the representations of the Fermions
of the self-dual string world sheet (\ref{sdfermionreps}), theory has chiral Fermions
and so possess anomalies. Using the results of \cite{alvarezgaumewitten} we
may read off the overall resulting normal bundle anomaly. (Also see
\cite{harvey2} for a good review of these issues). The normal bundle of
the string splits up into two SO(4) bundles, one that is tangent to the
five-brane and one that is normal to the five-brane. The normal bundle tangent
to the five-brane we denote by T and the normal bundle that is normal
to the five-brane we denote by N.

The anomalies are given through descent \cite{alvarezgaumewitten} by:
\beq
I_4=\pi (\chi(T) + \chi(N))
\eeq
where $\chi(T/N)$ denotes the Euler character of the SO(4) T/N
bundles respectively.
Given theory is anomalous we must introduce local terms in the
five-brane world volume and the self-dual string that will cancel this anomaly.

To cancel the T bundle anomaly one has a term:
\beq
I_{mc}=Q_2 \int_{M^6} b_2 \wedge \delta(\Sigma_2 \hookrightarrow M^6)
\eeq
on the five-brane world volume. $\delta(\Sigma_2 \hookrightarrow M^6)$
denotes the Poincare dual of the string world volume $\Sigma_2$ in the
five-brane world volume $M^6$. This is just the minimal
coupling of the string to the two form on the five-brane under which is it charged.
Its variation under SO(4) T-bundle transformations is:
\beq 
\delta I_{mc}= Q_{SD} \pi \int_{\Sigma_2} \chi_2^{(1)}
\eeq
where we have used the representation of the Poincare dual to be:
\beq
\delta_4(\Sigma_2 \hookrightarrow M^6) = d \rho(r) \wedge e_3/2 \, .
\eeq
with $\rho(r)$ the bump form and $e_3$ the global angular form over
the $S^3$ fibres transverse to $\Sigma_2$ in $M^6$ \cite{bermanharvey}.
Note the minimal coupling is proportional to $Q_{SD}$ and so this
indicates the world volume theory for an arbitrary number of strings
has an anomaly that scales linearly with $Q_{SD}$. This agrees with
the absorption cross section described in the previous section.

The anomaly in the N bundle may be cancelled by a term that originates
from the term introduced by
Ganor and Motl \cite{GanorMotl} and Intriligator \cite{Intrill} to cancel
anomalies of five-branes in the Coulomb branch. That is the
Wess-Zumino described in (\ref{WZ2}).
Their term, for the case of a self-dual string, where one excites a single scalar reduces to:
\beq
I= {{-1 } \over 2} \alpha(Q_5)  \int_{M^6} H_3 \wedge \chi^{(0)}(A_N)
\eeq
which is the term required to cancel the N bundle anomaly of the
self-dual string.
The crucial part is to examine the charge dependence of this term. It
is linear in $H_3$ which means it is linear in $Q_{SD}$.
The $Q_5$ dependence is given by (\ref{alpha}).
For the case $G=SU(Q_5+1), H=SU(Q_5)$ then
$\alpha={1 \over 2} Q_5$ which gives a linear dependence. The more
surprising result is if $G=SU(Q_5+1), H=U(1)^{Q_5}$ then $\alpha={1
  \over 8}(Q_5^2+Q_5-1) $. This implies the anomaly is proportional to 
\beq
Q_{SD} Q_5^2   \, ,
\eeq
in the large $Q_5$ limit. This $Q_5^2$ dependence is something that 
cannot be seen at this point
by any other means.\cite{alvarezgaumewitten})
Here we have discovered a further M-theory mystery, what are these
$Q_5^2$ degrees of freedom on the self-dual string? 

How might this be related to the degrees of freedom on the
five-brane? Well if one considers the case of maximal breaking where
$H=U(1)^{Q_5-1}$ then we are left with $Q_5-1$ tensor multiplets and each tensor
multiplet may have a self-dual string associated with it. Each string
will carry  $Q_5^2$ degrees of freedom so that in total all the
self-dual strings will carry $Q_5^3$ degrees of freedom which agrees
with the five-brane in the large $Q_5$ limit. Note, however the next to
leading order in $Q_5$ does not agree; for the five-brane there  is no
term quadratic in $Q_5$ whereas there is such a dependence if one
counts self-dual string degrees of freedom. Work on anomalies of the
self-dual string has also been discussed in \cite{french,mansanom}.

\subsection{AdS limits of the self-dual string}

One may try to construct a decoupling limit in which the low energy
dynamics of the self-dual
string decouple from the five-brane bulk \cite{berman4}. 
This will be analogous to
the Maldacena near horizon limits for branes.

One takes a low energy limit while keeping fixed some energy scale:
\beq
l_p \rightarrow 0 \qquad u={r^2 \over l_p^3} \,  {\rm{fixed}}  \, .
\eeq

This limit is determined simply by the dimensions of the two
dimensional string scalars or via appealing to the description in
terms of membranes. This is idententical to the Maldacena limit for
membranes (as one might expect).
Taking this limit for $Q_{SD}$ fixed, gives the following. (Note,
$Q_{SD}>>1$ so that the five-brane equations of motion are valid, ie. there
are no derivative corrections). The effective  metric (\ref{OMmetric}) which
governs the dynamics of the brane becomes conformal to: $AdS_3 \times S^3$. The
radius of the $AdS_3$ is given by:
\beq
R_{AdS_3}= R_{S^3} = ({2} Q_{SD})^{1 \over 3} l_p  \, .
\eeq
This is of course for $Q_5=1$. The area of the sphere is therefore
proportional to $Q_{SD}$, in keeping with our intuition
that the cross section goes linearly with $Q_{SD}$.
We cannot repeat the calculation for the case where the
$Q_5>1$ since we don't have a description of the five-brane theory in
this case.
 
In the above limit the effective tension of open membranes is constant in units of
$l_p$ and so the open membranes do not decouple \cite{OM}. The bulk five-brane theory is
therefore described by open membranes on an $AdS_3 \times S^3$ background with
a decoupling of the closed membrane sector. Therefore the low energy
dynamics of the decoupled
self-dual string (in the large $Q_{SD}$ limit) appears to be described
by open membranes on $AdS_3 \times S^3$. Unfortunately we do not have
sufficient control of either side of this correspondence to make
further statements.

\subsection{Self-dual string effective actions}

A conjectured action for a single self-dual string has been given by
\cite{henningson,hen2,hen3,hen4}. Of course the self-dual string as described above 
has infinite tension. To get a finite tension string one
imagines the open membrane ending on another five-brane so the string
has a finite effective tension. The low energy effective action of the
Bosonic sector is
given by (where low energy means small with respect to the energy scale
given by the inverse of the five-brane separation):
\beq
S= {1 \over {l_p^3}} \int d^2 \sigma  |\phi| ( \sqrt{{{\rm{detg}}_{\mu \nu}}}
+ d X^M \wedge d X^N b_{MN} )  \, ,
\eeq
where $g_{\mu \nu}$ is the induced world sheet metric. The first term
is essentially a Nambu-Goto term but with a tension set by
$|\phi|$. The second term is just the minimal coupling to the
five-brane world volume two form b. 
One may check that this captures the dynamics described above. The scattering
amplitude of waves off the string has been calculated in
\cite{henningson}. Comparison with \cite{berman4} agrees with the
scattering off the self-dual string soliton.

Further work has been done with this action. However, it has a
restricted range of validity and is probably not valid when there are
more than one coincident strings or five-branes. One further approach may
be to calculate the moduli space of two (or more) strings and
determine their effective action as a sigma model on that moduli space. This
is just what one would do for the effective action of two monopoles
for example.

\section{The membrane boundary theory}

In the preceeding sections the membrane has been described as a
five-brane soliton - the self dual string and its dynamics analysed as a
low energy effective description for the soliton. We now move to
describe the open membrane directly by using the action for the single membrane
and paying special attention to the minimal coupling of the C field to
the membrane.

\subsection{The boundary of the membrane}

\beq
\int_{M^3} C_{IJK} \pl_\mu X^I \pl_\nu X^J \pl_\rho X^K \epsilon^{\mu
\nu \rho}
\eeq
is the term that dictates the boundary membrane dynamics when there is
a large $C$ field. As such, the large C field limit provides an expansion
for the membrane dynamics. It is instructive to see how the open membranes
behave in this limit. We begin by choosing $C$ to be constant on the
five-brane. Since the pullback of $C$ is also governed by the nonlinear
self-duality relation  one needs to take the limit in a way that
preserves this equation of of motion. A description of the details of
this limit are given in \cite{bergshoeff1}. Here we will note
that the limit exists and analyse the boundary of the membrane in the
presence of a constant $C$ field given by $C_{IJK}= \epsilon_{IJK} C$ where I,J,K
are indices labelling a three dimensional subspace of the five-brane.

For such a constant $C$ field the above minimal coupling term
becomes:
\beq
S=\int_{\pl M^3} C \epsilon_{IJK} X^I \pl_\sigma X^J \pl_\tau X^K   \label{baction}
\eeq

This is a first order action. Its canonical momentum will be:
\beq
P_I= C \epsilon_{IJK} X^J \pl_\sigma X^K \, .
\eeq
This momentum relation must then be imposed as a constraint because the
action is first order and so the time derivatives will not be related to the momentum.
Calculating the brackets of these constraints implies that they
are second class which in turn implies one should replace Poisson brackets with Dirac brackets.  
We then calculate the Dirac bracket for the $X^I(\sigma)$ \cite{bergshoeff1}:
\beq
[X^I(\sigma), X^J(\sigma')]_D = \delta(\sigma -\sigma') {{\epsilon_{IJK} \pl X^K (\sigma)} \over {  |\pl
    X| C }}   \label{ncloop}
\eeq
Now we see that the $X^I(\sigma)$ do not commute. This is the M-theory
analogue of noncommutativity on a D-brane in the presence of a Neveu
Schwarz two form. In that case, the boundary of a string was a point
and the noncommutativity was an ordinary sort. For example, for a
constant Neveu-Schwarz two from in a two dimensional subspace of the
D-brane $B_{IJ}=\epsilon_{IJ} B$ a similar analysis for the open
string in the large B limit yields:
\beq
[X^I,X^J]= {{\epsilon^{IJ}} \over B} \, . \label{ncplane}
\eeq
The M-theory relation (\ref{ncloop}) is therefore interpreted as defining
a noncommutative loop space in comparison to the simple noncommutative space described by (\ref{ncplane}). 
Physically this occurs is because the membrane boundary is a loop as
opposed to the case of the open
string which has a point-like boundary.

How one interprets this from the five-brane perspective is difficult
and we will return to this later.

\subsection{From strings to ribbons}
We will now consider the case where the
membrane has two boundaries on the same 5-brane
in the presence of non-vanishing flux. 

First we observe that \cite{boris} for a single boundary the equations of motion of 
the boundary string
\be
C_{ijk} ~ \pa_\sigma X^j \wedge \pa_\tau X^k = 0 
\ee
imply that $X^i$ has to depend on the  worldsheet coordinates
through only one function $f(\tau,\sigma)$. The boundary of the
membrane therefore  is a  static string in the $(X^1,X^2,X^3)$
plane. A useful analogy is that of a vortex
line in a three-dimensional  fluid: just as in two-dimensions, 
the transverse motion  of a vortex line  is
effectively confined by the rotational motion of the fluid itself,
on Landau-like orbits. In addition, there exist soft  modes
propagating along the vortex line  known as Kelvin modes. In fact,
as noticed  in  \cite{Matsuo:2001fh,boris}, 
the boundary coupling (\ref{baction}) is precisely
the one describing  the Magnus effect in fluid hydrodynamics.

Membranes with two boundaries, however, have no overall
charge and therefore can propagate freely just as vortex anti-vortex
lines may propagate in hydrodynamics.  In the
absence of a $C$ field,  the two boundaries
lie on top of each other, leading to an effectively tensionless string,
the tentative fundamental degrees of freedom
of the (0,2) theory.  In the presence of a $C$ field however, these
tensionless strings polarise into thin
ribbons, whose width is proportional to the local momentum density.
Indeed, the canonical momentum on the membrane, neglecting the
contribution of the Nambu-Goto term, is
\be
P^i =C_{ijk} ~\partial_\sigma X^j \partial_\rho X^k
\ee
where $\sigma$ is the coordinate along the boundary string, and
$\rho$ the coordinate normal to it. The ribbon thus grows as
\be
\Delta^i \sim \partial_\rho X^i= 
\frac{1}{C |\pa_\sigma X|^2} \eps_{ijk} P^j \pa_\sigma X^k 
\ee
where we retain in $\Delta$ only the component orthogonal to 
$\sigma$ (the parallel component could be reabsorbed by
a diffeomorphism on the membrane worldvolume).

Let us consider a
simple classical solution corresponding to an infinite strip of width
$\Delta$ moving at a constant velocity $v$ transverse to it: We thus
consider the classical solution 
\be 
\label{zmm}
X^i = p_0^i \tau + u^i \sigma + \Delta^i \rho \; \; . 
\ee 
The boundary condition 
\be  \sqrt{\gamma}
\gamma^{\rho\rho} \pa_\rho X^i - C_{ijk} \pa_\sigma X^j \pa_\tau X^k
= 0 
\ee 
with induced metric $\gamma=\diag(m^2,|u|^2,|\Delta|^2)$, implies
that the direction of the polarisation vector is orthogonal to the plane
formed by the tangent vector to the string $\vec u=\partial_\sigma \vec X$
and the local velocity $\vec p_0 = \partial_\tau \vec X$, 
\be
\frac{\vec \Delta}{|\Delta|} = C \frac{\vec u}{|u|} \wedge \frac{\vec
p_0}{m} \, .
\ee 
Calculating the local canonical momentum
\bea 
\label{cm}
P^i &=& 
\sqrt{\gamma} \gamma^{\tau\tau} \pa_\tau X^i - C_{ijk} ~ \pa_\sigma
X^j ~ \pa_\rho X^k \\ &=&
\frac{|\vec u|~|\vec\Delta|}{m(1+C^2)} ~\left[ ( 1+ C ^2) p_0^i -
C^2 \frac{\vec u \cdot \vec p_0}{|\vec u|^2}  u^i \right] 
\eea 
one may express the local velocity in terms of $P^i$,
\be
\label{poi}
p_0^i = \frac{m}{|\vec u|~|\vec\Delta|(1+C^2)}
\left[ P_i + C^2 \frac{\vec u \cdot \vec P}{|\vec u|^2} u^i \right] 
\ee
and obtain the relationship between the membrane polarisation and 
canonical momentum,
\be \vec \Delta = \Theta \frac{\vec
u}{|u|^2} \wedge \vec P\ \ , \qquad \Theta = \frac{C}{1+C^2} \, .
\label{th}
\ee
For convenience, we will use the gauge $m=|\vec u|~|\vec \Delta|$
from now on.

We thus recover the ``open membrane non-commutativity parameter'' $\Theta$,
defined in \cite{Berman:2001rk}. In this work, this parameter was determined by studying
the physics of five-branes probing
supergravity duals with $C$-flux longitudinal to the probe brane world
volume. We now understand this result classically as the 
polarisability of open membranes in a $C$-field. Of course in the
large $C$ limit it becomes $1 \over C$ as we expect from (\ref{ncloop}) and
is the obvious generalisation to the open string non-commutativity
parameter \cite{bermanpioline}.

\subsection{An effective Schild action for string ribbons}

We will now describe an effective string theory that describes the
ribbons as strings. This is a good approximation for scales
less than the ribbon width \cite{bermanpioline}.
Let us start with the light-cone formulation \cite{hoppe} 
of the membrane, with Hamiltonian
\be
\label{hop}
P^- = \int d\sigma d\rho ~\frac{1}{2P^+}
\left[ (p_0^i)^2  +  g \right]
\ee
where $g$ is the determinant of the spatial metric, hence the square
of the area element (and the membrane tension is set to 1). 
In this gauge, one should enforce the constraint 
\be
\pa_\sigma X^i \pa_\rho \pa_\tau X^i - 
\pa_\rho X^i \pa_\sigma \pa_\tau X^i =0  \, ,
\ee
which is trivially satisfied on zero-mode configurations (\ref{zmm}).
For a thin ribbon of width $\vec\Delta$ given
by (\ref{th}), the square of the area element is
\be
g=| \vec u \wedge \vec \Delta|^2 =
\frac{C^2}{(1+C^2)^2} \left[ \vec P^2 - \frac{(\vec u \cdot \vec P)^2}
{|u|^2} \right] \, .
\ee
On the other hand, using (\ref{poi}), the kinetic energy may be written as
\be
(\vec p_0)^2 = \frac{1}{(1+C^2)^2} 
\left[ \vec P^2 + C^2 (C^2+2) \frac{(\vec u \cdot \vec P)^2}{|u|^2}
\right] \, .
\ee
Finally, the total Hamiltonian thus takes the form
\be
\label{h1}
P^- = \int d\sigma \frac{1}{2P^+ (1+C^2)} 
\left[ P^2 + C^2 \frac{(\vec P \cdot 
\pa_\sigma \vec X)^2}{| \pa_\sigma X^i|^2} \right] \, .
\ee
From this expression, specifying to a gauge choice where $\vec P$
and $\pa_\sigma \vec X$ are orthogonal, we see that the effective
metric in the transverse directions is rescaled by a factor of
$(1+C^2)$,
\be
G_{ij} = \left[ 1 + C^2 \right] \delta_{ij} \, .
\ee
This agrees with the membrane metric found from very different considerations
in \cite{Berman:2001rk,VanderSchaar:2001ay}, up to the conformal factor  $Z=(1-\sqrt{1-1/K^2})^{1/3}$ 
with $K=\sqrt{1+C^2}$. 

Finally, we may perform a Legendre transform on $P_i$
to find the Lagrangian density of the ribbon,
\be
\label{lag}
{\cal L}=\int d\sigma ~\frac12 (\pa_\tau X^i)^2 
+ \frac{C^2}{2| \pa_\sigma X^i|^2} \sum_{i,j} 
\{ X^i,X^j \}^2
\ee
where we have defined the Poisson bracket on the Lorentzian string 
worldsheet\footnote{This should not be confused with the Poisson bracket
formulation of the membrane, which refers to the two {\it spatial} directions
of the membrane world-volume.} as $\{A,B\}=\pa_\sigma A \pa_\tau B - \pa_\sigma B \pa_\tau A$. Note that 
the relative sign between the two terms in (\ref{lag}) is consistent
with the fact that they both contribute to kinetic energy. For vanishing
$C$, (\ref{lag}) reduces to the Lagrangian for a tensionless string,
as expected. While we have mostly worked at the level of zero-modes,
it is easy to see that (\ref{lag}) remains correct for arbitrary
profiles $X^i(\tau,\sigma)$, as long as the dependence on 
membrane coordinate $\rho$ is fixed by Eqs. (\ref{zmm}), (\ref{th}).

After fixing the invariance
of the Lagrangian (\ref{lag}) under general reparameterizations of $\sigma$
by choosing $| \pa_\sigma X^i|=1$, we recognise in the second term the Schild 
action, which provides (in the case of a  Lorentzian target-space)
a unified description of both tensile and tensionless strings, 
depending on the chosen value for the conserved quantity 
$\omega=\{ X^i , X^j \}^2$~\cite{Schild:1976vq}. This term dominates over the first
in the limit of large $C$ field. For any finite value however, $\omega$ is
not conserved, and the second term in (\ref{lag}) can be interpreted as
the action for a non-relativistic string with tension proportional to $C$.

As usual, it is possible to give a regularisation of this membrane
action, by replacing the Poisson bracket (now in light-cone directions
on the worldsheet) by commutators in a large $N$ matrix model.
One thus obtains a lower-dimensional analogue of the type IIB IKKT
matrix model \cite{Ishibashi:1996xs},
\be
P^- = \frac{1}{2P^+} \left(
[A_0,X^i]^2 
+  C^2  \sum_{i<j} [ X^i, X^j ] ^2 \right)\, .
\ee
The analysis of this matrix model has not been carried out. It would
be interesting to see what one could learn about membrane ribbons from
quantising this action.

\subsection{Non-commutative string field theory}

Just like open strings, open membranes interact only when their
ends coincide. Since their  boundaries  are tensionless
closed strings which polarise into thin
ribbons in the presence of a strong $C$ field, one may expect
that the effect  of  the $C$  field can be encoded by a deformation
of the  string field theory describing the membranes boundaries.
Despite the fact that string  field theory  of closed  strings, 
not to mention tensionless ones, is a rather ill-defined subject,
it is natural to represent the string field as a functional 
in the space of loops. The effect of the polarisation of
the ribbons can thus represented  by
\bea
\label{vstar}
V &\sim& \int [DX(i)] 
\Phi\left[ X^i- \frac12 \frac{\Theta}
{ |\pa_\sigma X|^2} \eps_{ijk} \pa_\sigma X^j 
\frac{\delta}{\delta X^k}\right]~ \nonumber\\ &&\times
\Phi\left[ X^i+ \frac12 \frac{\Theta}
{ |\pa_\sigma X|^2} \eps_{ijk} \pa_\sigma X^j 
\frac{\delta}{\delta X^k}\right]
\eea
where we represented  the momentum density $P_i$, canonically
conjugate to $X^i(\sigma)$, as a derivative operator in the space of loops.
Defining the operators
\be
\tilde X^i(\sigma) =  X^i- \frac12 \frac{\Theta}{ |\pa_\sigma X|^2}
\eps_{ijk} \pa_\sigma X^j \frac{\delta}{\delta X^k}
\ee
reproduces the non-commutative loop space in the
``static'' gauge $X^3(\tau,\sigma)=\sigma$,
\be
[ \tilde X^1(\sigma), \tilde X^2(\sigma') ]
= \Theta \delta(\sigma-\sigma')
\ee
as proposed in \cite{bergshoeff1,Kawamoto:2000zt}. 
The fact that the transverse fluctuations of a vortex line
are effectively confined by an harmonic potential is well
known in fluid dynamics. In the more covariant
gauge $|\pa_\sigma \vec X|=1$, one  obtains a
tensionless limit of the $SU(2)$ current algebra,
\be
[ \tilde X^i(\sigma), \tilde X^j(\sigma') ]
= \Theta \eps_{ijk} \pa_\sigma X^k \delta(\sigma-\sigma')  \, .
\ee
The same relations may be directly obtained by Dirac quantisation
of the topological open membrane Lagrangian (\ref{baction}).

More generally, much as  in the non-commutative
case, this deformation amounts to multiplying the closed string
scattering amplitudes by a phase factor proportional to the
volume enclosed  by the ribbons as  they interact. 

In this section we have seen how the presence of a C-field on
the five-brane deforms the membrane boundary and its canonical
quantisation  leads to a noncommutative loop space on the
five-brane. This also could be viewed as a thickening of the membrane
boundary into ribbons. Recently there was a proposal for an effective
description in terms of a field theory with nonassociative algebra \cite{omnaqft}.
The nonassociativity arises from a product made with the ribbons just
as a product of open strings leads to a noncommutative field theory.

There are many open questions on key aspects of the quantum
nature of this nonassociative theory. One hint, however, that this
may be in the right direction is the appearance of nonassociativity in
the membrane five-brane system from other considerations which will
become apparent later.

\section{Five-Branes from membranes}

We now move to describing the membrane ending on the five-brane from the
membrane perspective. The idea is that this will be a BPS solution of some multiple membrane
effective action. This is in analogy with the description of the D1 brane ending on a D3
brane  where, from the D1 brane perspective, the system is
described as a {\it{fuzzy funnel}} solution of the D1 brane ${1 \over
  2}$ BPS equation \cite{CMT}. Essentially the D1 has a (fuzzy) two
sphere cross section whose radius diverges to give an additional three
dimensions forming the D3 brane. The diverging fuzzy two sphere is known as
a fuzzy funnel. 

The goal will be to reverse engineer the membrane action. First 
we decide on the properties of the required solution needed to describe
a membrane ending on a five-brane. Then we construct an
equation whose solutions have such properties and finally we will be
able to deduce
the action from which that equation is a Bogmolnyi equation. 
It should be stressed that this is all conjectural since the starting 
point cannot be derived from some fundamental action as
the non-Abelian D-brane action can. Instead the justification will that
the equation has the right properties required to describe the
membrane five-brane system. This is essentially M-theory
phenomenology. We will follow thesis by Neil Copland
\cite{coplandthesis} and the original work of Basu and Harvey, \cite{harvey1}.

\subsection{Expected properties for the membrane fuzzy funnel}{\label{prop}}

We know that the five-brane picture of the membrane ending is in terms of the
self-dual string. Examining the self-dual solution (\ref{sdstringsolution}) we
see the relationship between the radial direction, $R$, tangent to the five-brane
(i.e. $R=\sqrt{(X^1)^2+(X^2)^2+(X^3)^2+(X^4)^2}\,$), and $s$, the
membrane worldvolume direction away from the five-brane, will be: 
\beq\label{sdrad}
s\sim \frac{Q}{R^2},
\eeq
(Q is the membrane charge). As the self-dual string is a static
solution with no dependence on $\sigma$, 
the coordinate along the string, the active scalars should only depend on $s$.

In the D1-D3 system there were three active scalars transverse to the
string in the directions of the D3-brane worldvolume. 
This meant the cross section was a (fuzzy) two-sphere, as expected for
a BIon spike. In the membrane five-brane system we have an extra
transverse dimension so the cross section 
should be a (fuzzy) three-sphere, associated to the $SO(4)_T$ R-symmetry of the self-dual string.

Coordinates on a fuzzy three sphere, $G^i$, obey the equation
\beq\label{FS3}
G^i +\frac{1}{2(n+2)}\epsilon_{ijkl}G_5G^jG^kG^l =0,
\eeq
where $G_5$ is a constant matrix obeying $\{G_5,G^i\}=0$ and
$(G_5)^2=1$, \cite{harvey1}. 
$n$ is associated to an SO(4) representation as follows.
Matrices in the fuzzy sphere algebra may be thought of in terms of
SO(4) representations; SO(4) is decomposed into $SU(2) \oplus SU(2)$
and under this decomposition the fuzzy three sphere matrices are
restricted to be in $( {{n+1} \over 4},{{n-1} \over 4} )$ with n
labelling the representation. (In what follows we move between using N
as referring to $N \times N$ matrices and n as referring to the above
SO(4) representation associated to the fuzzy sphere algebra.)
This equation
is a quantised version of a higher Poisson bracket equation for a
three sphere and was first derived in \cite{harvey1}.

\subsection{The Basu-Harvey Equation}

By analogy with the D1-D3 system the Basu-Harvey equation is
conjectured to be a Bogomol'nyi equation for minimising the energy. 
It should also, as usual, follow from the vanishing of the 
supersymmetry variation of the Fermions on the membrane.

The Basu-Harvey equation is given by 
\beq\label{BaH}
\frac{dX^i}{d s}+\frac{M_{11}^3}{8\pi\sqrt{2N}}\frac{1}{4!}\epsilon_{ijkl}[G_5,X^j,X^k,X^l]=0.
\eeq
The anti-symmetric 4-bracket is a sum over permutations with sign, e.g.
\beq\label{N4b}
[X^1,X^2,X^3,X^4]=\sum_{\mbox{perms}\ \sigma} \mbox{sign}(\sigma)X^{\sigma(1)}X^{\sigma(2)}X^{\sigma(3)}X^{\sigma(4)}\, ,
\eeq
and it can be thought of as a quantum Nambu bracket\cite{Nambu,CZ}. We
could have used such a bracket for the 
second term of (\ref{FS3}) once the correct
combinatorial factor is included. 

$G_5$ and the scalars will belong to an algebra containing the fuzzy
three-sphere. 
There are three main possibilities for what this algebra is. 
The first is \mnc, the algebra of $N \times N$ matrices, where $N$ is
the dimension of the representation of $SO(4)$.  
For the fuzzy three-sphere only, this dimension coincides with the
square of the radius in terms of the $G^i$ (i.e. $\sum_i tr G^iG^i=N$). 
$N$ is what we will identify with the number of membranes, and this is
the $N$ that appears in the Basu-Harvey equation (\ref{BaH}). A second
possibility for the algebra is that 
generated by the $\{G^i\}$, which is a sub-algebra of \mnc. 
The third option is the algebra which in the large-$N$ limit agrees
with classical algebra of functions on $S^3$, namely the spherical
harmonics. For now, we will assume our fields are in \mncs.

\subsection{The Membrane Fuzzy Funnel Solution}
We expect a static solution with scalars
proportional to the fuzzy three sphere coordinates $G^i$ and only depending on $s$. An ansatz 
\beq
X^i(s)=f(s)G^i
\eeq
leads quickly to the solution
\beq\label{soln}
X^i(s)=\frac{i\sqrt{2\pi}}{M_{11}^\frac{3}{2}}\frac{1}{\sqrt{s}}G^i.
\eeq
The physical radius is given by
\beq\label{traces2}
R=\sqrt{\left|\frac{\mbox{Tr} \sum(X^i)^2}{\mbox{Tr} \openone}\right|}.
\eeq
This implies,
\beq
s\sim \frac{N}{R^2} \, ,
\eeq
which is the self dual string behaviour we expect when we identify $N$ with the number of membranes.
Other solutions to the Basu-Harvey equation have been found that
describe multiple parallel five-branes \cite{norgadi}. Later we will
describe generalisations of the equation to allow the description of
arbitrary five-brane calibrated geometries.

\section{An Action for Multiple Membranes}\label{enact}

Given that the Basu-Harvey equation should arise as a Bogomol'nyi
equation, we define the energy of our static configuration with four
non-zero scalars to be\cite{harvey1}:
\bea \label{m2energy}
E = T_2 \int d^2 \sigma {\rm Tr} \bigg[ \left( \frac{d X^i}{ds} + 
\frac{M_{11}^3}{8\pi\sqrt{2N}}\frac{1}{4!} \epsilon_{ijkl} 
[G_5, X^j, X^k, X^l] \right)^2  \cr 
+ \left(1- \frac{M_{11}^3}{16\pi \sqrt{2N}} \frac{1}{4!}\epsilon_{ijkl} \left\{ \frac{d X^i}{ds} ,
[G_5, X^j, X^k, X^l] \right\} \right)^2 \bigg]^{1/2} .
\eea
The membrane tension $T_2$ is given by $T_2=M_{11}^3/(2\pi)^2$ and the
integral is over the two spatial worldvolume directions 
$\sigma$ and $s$. In what follows we will consider the $X^i$ to obey
$\{G_5,X^i\}=0$. 
In terms of the fuzzy three-sphere algebra this means restricting the
$X^i$ to lie in $Hom(\cRp,\cRm)$ or $Hom(\cRm,\cRp)$. 
This is of course obeyed by $\{ G^i \}$. 
It means that by multiplying out the squares in
(\ref{m2energy}) $G_5$ can be eliminated.

If the scalars obey the Basu-Harvey equation then Bogomol'nyi bound is satisfied and the energy density linearises:
\beq 
E = T_2 \int d^2 \sigma {\rm Tr} \left( 1- \frac{M_{11}^3}{8\pi \sqrt{2N}} \epsilon_{ijkl} 
\frac{d X^i}{ds} G_5 X^j X^k X^l\right).
\eeq
The energy can be written for large $N$ as
\beq\label{BHenergy}
 E = N T_2 L \int ds + T_5 L \int 2\pi^2 dR R^3,
\eeq
where we have used $T_5=M_{11}^6/(2\pi)^5$ and $L$ is the length of
the string. These two terms have the energy densities 
you would expect for $N$ membranes and a single five-brane respectively.

We would expect this analysis to be only valid at the core
($s\rightarrow\infty$), though for large $N$ 
it agrees with the M5-brane picture, a description which should only
be valid in the opposite limit. 
Examining (\ref{m2energy}) we can see that a Taylor expansion in terms of powers of $X^i$ is valid when $M_{11}^6R^6N^3\ll 1$, that is $R\ll \sqrt{N}M_{11}^{-1}$. Thus if $N$ is large, $R$ can be large as well.

Given the expression for the energy (\ref{m2energy}) we can expand and
deduce terms in the associated action. Also the action should be a
function of the eight transverse scalars and allow $\sigma$
dependence. This reasoning yields:
\bea \label{m2act} 
S &=& -T_2 \int d^3 \sigma {\rm Tr} \bigg[ 1 + 
\left( \p_a X^M \right)^2
-\frac{1}{2N.3!}[X^M,X^N,X^P][X^M,X^N,X^P]
\non\\
&+&\frac{1}{2N.4.3!} \left[ \p_a X^L ,[X^M,X^N,X^P] \right]  \times
\bigg( \left[\p^a X^L ,[X^M,X^N,X^P] \right] 
\non\\&+&\left[ \p^a X^M 
,[X^L,X^P,X^N] \right] + \left[ \p^a X^N ,[X^L,X^M,X^P] \right]\non\\ &+&\left[ 
\p^a X^P ,[X^L,X^N,X^M] \right] 
\bigg)  +\ldots \bigg]^{1/2} \eea
for a multiple membrane action. $L,M,\ldots$ labels the 8 transverse
directions and $a,b,\dots$ the 3 worldvolume directions. 
The three-bracket used is defined analogously to the quantum Nambu
4-bracket (\ref{N4b}) as a sum over the six permutation of the entries, with sign.

Just as in the case of a single membrane one may use supersymmetry to
limit the form of the action. Significant progress in this
direction has made in \cite{howelindstrom}.

\subsection{Fluctuations on the Funnel}

We now analyse the fluctuations analysis on the membrane fuzzy funnel
in the four directions transverse to both the membrane and the
five-brane. We take a general kinetic term and the sextic coupling
given in the last section and carry out a linear analysis of the membrane fluctuations.
In flat space and static gauge, the pull back of the metric is given
by $P[G]_{ab}=\eta_{ab}+\partial_aX^M\partial_bX^M$, taking the
determinant will lead to the first two terms of (\ref{m2act}).
This gives the action used for fluctuation analysis, 
\beq\label{fluctact}
S=-T_2 \int d^3\sigma\mbox{Tr} \sqrt{-\det(P[G]_{ab})-\frac{1}{2N}\frac{1}{3!}[X^M,X^N,X^P][X^M,X^N,X^P]}\, .
\eeq
The fluctuations may depend on all three worldvolume coordinates and are proportional to the identity in the fuzzy sphere algebra, $\delta X^m(t,s,\sigma)=f^m(t,s,\sigma) \openone_N$. Keeping terms up to quadratic order in the fluctuations, gives
\beq
[X^M,X^N,X^P][X^M,X^N,X^P]=3(f^m)^2[X^i,X^j]^2\, ,
\eeq
where $M,N,P$ run over all indices, $m$ runs over the directions
transverse to the both branes and 
$i,j$ run over the non-zero scalars of the solution. Evaluation of the commutator squared on the right-hand side proceeds via
\beq
[G^i, G^j]^2 = 2 G^i G^j G^i G^j -2 N^2\openone
\eeq
and
\beq\label{ijij}
G^i G^j G^i G^j\Prp=-(n+1)(n+3)\Prp=-2N\Prp.
\eeq
Finally this leads to 
\beq\label{ijsq}
[G^i, G^j]^2=-2N(N+2)\, .
\eeq
Returning to the action (\ref{fluctact}) we now have
\beq
S = -T_2 \int d^3 \sigma {\rm Tr} \sqrt{ H -H (\p_t f^m)^2 +(\p_s f^m)^2 
+H (\p_\sigma f^m)^2 +\frac{N+2}{2s^2} (f^m)^2  },
\eeq
where 
\beq
H=1+\frac{\pi N}{2M_{11}^3s^3}.
\eeq
The equation of motion for the linearised fluctuations becomes
\beq
(H \p_t^2 -\p_s^2 -H \p_\sigma^2 ) f^m (t,s,\sigma) 
+\frac{N+2}{2s^2} f^m (t,s,\sigma) =0.
\eeq
In the $s\rightarrow\infty$ limit (where we have a flat membrane) the equation of motion reduces to
\beq
(-\p_t^2 +\p_s^2 +\p_\sigma^2) f^m =0.
\eeq
The solutions to this equation are plane waves with $SO(2,1)$ symmetry in the worldvolume directions, as one would expect for a membrane.
Although in the opposite limit the analysis should not be valid, as per the earlier discussion we keep $N$ large and find agreement with what we would expect. As $s\rightarrow 0$, $H\sim s^{-3}$ and the equation of motion gives
\beq
(-\p_t^2 + \p_\sigma^2) f^m + R^{-3} \frac{\p }{\p R}
\left( R^3 \frac{\p f^m}{\p R} \right) =0.
\eeq 
This has exactly the $SO(2,1)\times SO(4)$ symmetry that we would expect for the M5 worldvolume with string soliton.

\section{Calibrations and a Generalised Basu-Harvey Equation}

While the Basu-Harvey equation is successful in reproducing many of
the properties desired for the M2-M5 system, 
it was not derived from any fundamental principle. Given this somewhat
ad-hoc construction, more detailed and involved checks are desirable to
establish its validity. The Basu-Harvey equation only excited four of the eight
scalars on the membrane. It is natural to consider whether the
Basu-Harvey equation may be generalised to include the other scalars
on the membrane. This will give rise to a membrane description of
five-brane calibrated geometries. Instead of the membranes blowing up into a
single five-brane we will describe membranes blowing up into five-brane
intersections or equivalently five-branes on calibrated cycles. (Recall
that a calibrated cycle is a minimal volume surface that
possesses a calibration form defining the cycle, see \cite{Harvey:1982xk} for
a description of calibrations.)
This is essentially an M-theory version of \cite{CL} where D1 branes
end on calibrated three brane geometries. This constitutes a good
check on the validity of the Basu-Harvey equation; the membranes have
to be able to describe any allowed supersymmetric five-brane configuration.

As in \cite{CL} we work with a `linearised' action to describe BPS
solutions of the coincident membrane theory. 
We then present the generalised Basu-Harvey equation, with the
conditions necessary for it to appear as a Bogomol'nyi equation. 
This is then encoded as a supersymmetry variation (though not actually
with a supersymmetric action). 
This generalised Basu-Harvey equation will have solutions that describe coincident membranes ending on 
intersections of five-branes corresponding to calibrated geometries. 
We will then list these possible configurations and some simple solutions
to give a flavour of the intricacies involved. 
Along the way, we must solve the generalised form of the BPS equation
and some algebraic conditions on the brackets. This leads to
additional algebraic conditions 
whose two-bracket equivalents were identically solved in the simpler
D1-D3 case. These algebraic conditions hint at the possibility of a
new algebraic structure for the membrane fields.

\subsection{A Linear Action for Coincident Membranes}

The generalised version of the Nahm equation was found as the
Bogomol'nyi equation of the {\it linear} action in \cite{CL}. 
The linear action is simply dimensionally reduced super
Yang-Mills. For BPS states, the linear action is in fact equal to the
full action (see \cite{Brecher} for a fuller discussion of this
point) this allows us to work with the linear action rather than the
much more complicated nonlinear Born-Infeld type action when dealing
with BPS states..
There is an analogous situation for the membrane theory. 
The starting point is the energy for coincident
membranes ending on a single five-brane (\ref{m2energy}).
If the Basu-Harvey equation (\ref{BaH}) holds 
and $\{G_5,X^i\}=0$, then the remaining terms under the square-root  can be rewritten as the perfect square
\beq
E=T_2\int d^2\sigma \mbox{Tr}\left[\left(1+\frac{1}{2}(\partial_\sigma
X^i)^2-\frac{1}{2N}\frac{1}{2.3!}[X^i,X^j,X^k]^2\right)^2\right]^{1/2} .
\eeq
This is the energy that one would get from an action
\beq\label{3action}
S=-T_2\int d^3\sigma
\mbox{Tr}\left(1+\frac{1}{2}(\partial_a
X^i)^2-\frac{1}{2N}\frac{1}{2.3!}[X^j,X^k,X^l]^2\right) \, ,
\eeq
which is the {\it{linearised}} form of the membrane action
(\ref{m2act}) for three non-zero scalars. 
This is the action we will use when looking for the generalised
Basu-Harvey equation but with the indices $i,j,k$ running over all
eight scalars. Since we are dealing with static configurations it is
sufficient to consider the energy functional ie. the Hamiltonian. 

As is usual in M-theory, we do not have a coupling constant in which
to expand.  We can however expand in powers of $X^i$ and it is in this
expansion that we are working to leading order.

\subsection{A Generalised Basu-Harvey Equation}

Consider the Hamiltonian,
 \beq
E=\frac{T_2}{2}\int d^2\sigma \mbox{Tr}\left( X^{i^ \prime}
X^{i^\prime }-\frac{1}{3!}[X^j,X^k,X^l][X^j,X^k,X^l]\right)
\eeq
(a trivial constant piece corresponding to the flat brane has been
subtracted). 
The indices $i,j,\dots$ run from 2 to 9 and $X^{10}$ is identified
with $\sigma$. The factor of $1/(2N)$, has been scaled out as was 
previously done with the numerical factors to simplify the
presentation. Reintroduction of this factor just involves 
inserting a factor of $1/\sqrt{2N}$ with each three- or four-bracket. 
We proceed by using the usual Bogomol'nyi construction to write 
\beq\label{m2bog}
E=\frac{T_2}{2}\int d^2\sigma \left\{\mbox{Tr}\left(
X^{i^\prime}+g_{ijkl}\frac{1}{4!}[H^*,X^j,X^k,X^l]\right)^2+T\right\}\, ,
\eeq
where $T$ is a {\it{topological}} piece given by
\beq T=-T_2\int
d^2\sigma
\mbox{Tr}\left(g_{ijkl}X^{i^\prime}\frac{1}{4!}[H^*,X^j,X^k,X^l]\right)\, .
\eeq
When there are only four non-zero scalars ($X^2,\ldots,X^5$) and
$g_{ijkl}=\epsilon_{ijkl}$ 
this gives the Basu-Harvey equation and the topological piece gives the energy of the five-brane on which the
membranes end. For more scalars $g_{ijkl}$ is essentially the
calibration form of the five brane on which the membrane ends.

If more than four scars are non-zero, then we must impose
\bea\label{fullconst}
&&\frac{1}{3!}g_{ijkl}g_{ipqr}\mbox{Tr}\left([H^*,X^j,X^k,X^l][H^*,X^p,X^q,X^r]\right)\\
\nonumber &&=\mbox{Tr}\left([H^*,X^i,X^j,X^k][H^*,X^i,X^j,X^k]\right) \eea
in order to be able to rewrite the action as in (\ref{m2bog}). $H^*$ is a more general form of $G_5$, chosen to have the analogous properties $\{H^*,X^i\}=0$ and $(H^*)^2=1$. In fact using these properties (\ref{fullconst}) reduces to the simpler form
\beq\label{mconstraint}
\frac{1}{3!}g_{ijkl}g_{ipqr}\mbox{Tr}\left([X^j,X^k,X^l][X^p,X^q,X^r]\right)=\mbox{Tr}\left([X^i,X^j,X^k][X^i,X^j,X^k]\right)
\, ,   
\eeq 
which is the M-theory version of the constraints given in \cite{CL}
for the D1-D3 brane system.

Once we have written the energy in the form (\ref{m2bog}) using (\ref{mconstraint}) then we can clearly minimise it by imposing the generalisation of the Basu-Harvey equation 
\beq\label{myNahm}
\pd{X^i}{s}+\frac{M_{11}^3}{\sqrt{2N}8\pi}g_{ijkl}\frac{1}{4!}[H^*,X^j,X^k,X^l]=0.
\eeq
This equation has factors restored, and $g$ is a general anti-symmetric four-tensor. When $g_{ijkl}=\epsilon_{ijkl}$ we recover the Basu-Harvey equation and (\ref{mconstraint}) is an identity. 

\subsection{The Equation of Motion}

The equation of motion following from the action (\ref{3action}) is given by 
\beq
X^{i^{\prime\prime}}=-\frac{1}{2}\lt X^j,X^k,[X^i,X^j,X^k]\rt
\eeq
where the three bracket $\lt A,B,C \rt$ is the sum of the six
permutations of the three entries, but with the sign of the permutation
determined only by the order of the first two entries; i.e. $ABC, ACB
\mbox{ and } CAB$ are the positive permutations.  By using the
Bogomol'nyi equation (\ref{myNahm}) twice on the left-hand side it is equivalent to: 
\beq
\label{3bconst} \frac{1}{3!} g_{ijkl} g_{jpqr}\lt
X^k,X^l,[X^p,X^q,X^r]\rt=-\lt X^j,X^k,[X^i,X^j,X^k]\rt \, \, .  
\eeq
After multiplying by $X^i$ and taking the trace, we recover the
constraint equation (\ref{mconstraint}). Thus in summary, the
solutions of the generalised Basu-Harvey equation (\ref{myNahm}) that
obey the algebraic equation of motion (\ref{3bconst}) are minimal energy solutions to the
equations of motion of the proposed membrane action (\ref{3action}).

\subsection{Supersymmetry}

In the D1-D3 system the Nahm equation could be derived {\it either} as
the Bogomol'nyi equation for minimising the energy, {\it or} as a
requirement for preserving half the supersymmetry.

Here we do not have a supersymmetry variation or indeed an action from which to start.
What we can do is to impose by fiat a simple generalisation of the
linearised supersymmetry variation for the D1-strings and determine
whether it leads to a consistent 
picture of membranes ending on five-branes. 

We find that if the generalised form of the
Bogomol'nyi equation is satisfied then it leads to a simplified form 
of the supersymmetry variation, where the route to further 
simplification is the imposition of a set of projectors corresponding
to the non-zero components of $g_{ijkl}$. Compatible sets of these
projectors are in correspondence with the known calibrated five-brane 
intersections and the imposition of these projectors leads to the
preservation of a certain fraction of supersymmetry, the fraction 
being that preserved by the corresponding five-brane intersection with
a membrane attached. 
This is of course provided we satisfy the algebraic conditions on the 
brackets left over in the supersymmetry condition after 
imposition of the projectors. 
Similar to the case of D3-brane intersections, 
the total space of all the intersecting five-branes 
in the submanifold calibrated by $g=\frac{1}{4!}g_{ijkl}dx^i\wedge dx^j\wedge dx^k\wedge dx^l$.

It remains to check if the algebraic conditions on the brackets are
enough to satisfy 
the constraint (\ref{mconstraint}) and thus the equations of motion. 
It turns out that it is not quite enough, there are additional
algebraic conditions, a set of equations of similar form for all
configurations, which must be satisfied to solve the constraint. 
For the D1-D3 case these had a simpler form and were satisfied identically. 

The most obvious suggestion for the supersymmetry variation is 
\beq
\delta\lambda=\left(\frac{1}{2}\partial_\mu
X^i\Gamma^{\mu
i}-\frac{1}{2.4!}\gtb{i}{j}{k}\Gamma^{ijk}\right)\epsilon. 
\eeq 
We substitute the generalised Basu-Harvey equation in the first term and
rearrange. The requirement that the supersymmetry variation
vanishes becomes that
\beq\label{susycond}
\sum_{i<j<k}\tb{i}{j}{k}\Gamma^{ijk}(1-g_{ijkl}\Gamma^{ijkl\#})\epsilon=0,
\eeq
where we have removed an overall factor of $H^*$ from the left-hand side
since, like $G_5$, it has trivial kernel. $\epsilon$ is the
preserved supersymmetry on the membrane world volume and we
have $\Gamma^{01\#}\epsilon=\epsilon$, where the membrane's worldvolume
is in the 0, 1 and $10=\#$ directions. We can then solve the
supersymmetry condition (\ref{susycond}) by defining projectors
\beq
\label{proj} P_{ijkl}=\frac{1}{2}(1-g_{ijkl}\Gamma^{ijkl\#})\, ,
\eeq
where there is no sum over $i,j,k \mbox{ or }l$. 

We normalise $g_{ijkl}=\pm1$ so they obey $P_{ijkl}P_{ijkl}=P_{ijkl}$. (Note, in all the
cases that we will consider, for each triplet $i,j,k$, $g_{ijkl}$ is only
non-zero for at most one value of $l$). We impose $P_{ijkl}\epsilon=0$
for each $i,j,k,l$ such that $g_{ijkl}\neq0$. Then by using
the membrane projection ($\Gamma^{01\#}\epsilon=\epsilon$) we can see
that each projector $P_{ijkl}$ corresponds to a five-brane in the
$0,1,i,j,k,l$ directions. To apply the projectors simultaneously, the matrices $\Gamma_{ijkl\#}$ need to commute with each other. $[\Gamma_{ijkl\#},\Gamma_{i'j'k'l'\#}]=0$ if and only if the
sets $\{i,j,k,l\}$ and  $\{i',j',k',l'\}$ have two or zero elements in
common, corresponding to five-branes intersecting over a three-brane
soliton or a string soliton.

Once we impose the set of mutually commuting projectors, the
supersymmetry transformation (\ref{susycond}) reduces to
\beq\label{susyids} \sum_{g_{ijkl}=
0}\tb{i}{j}{k}\Gamma^{ijk}\epsilon=0.  
\eeq
Here we sum over triplets $i,j,k,$ such that $g_{ijkl}=0$ for
all $l$. Using the projectors allows us to express these as a set of
conditions on the 3-brackets alone.

\section{Five-Brane Configurations}

We will now describe the specific equations that correspond 
to the various possible intersecting five-brane configurations. 

The five-branes must always have at least one spatial direction in common, 
corresponding to the direction in which the membrane ends. These configurations of five-branes can also be thought
of as a single five-brane stretched over a calibrated manifold \cite{Harvey:1982xk}. These
five-brane intersections can be found in 
\cite{GP,Jerome,Acharya:1998en}. We list the
conditions following from the modified Basu-Harvey equation, those following 
from the supersymmetry conditions
(\ref{susyids}) (with $\nu$ the fraction of preserved supersymmetry) and then 
discuss any remaining conditions required to
satisfy the constraint (\ref{3bconst}).

\subsection{Calibrated five-branes}

The first configuration is the original set up of a membrane ending on a single five-brane.
\bea\label{c1} 
g_{2345}=1\quad\quad\quad
\nu=1/2\quad {} \eea \bea {X^2}' = -H^*\tb{3}{4}{5}&,& \quad {X^3}' =
H^*\tb{4}{5}{2}\ , \quad \nonumber\\ {X^4}' = -H^*\tb{5}{2}{3}&,&
\quad {X^5}' = H^*\tb{2}{3}{4}\ . \nonumber\\ \nonumber \eea

The next case is two five-branes intersecting over a three-brane
corresponding to an 
SU(2) K\"{a}hler calibration of a two-surface embedded in four
dimensions. 
In terms of the first five-brane's worldvolume theory the condition
for preserved supersymmetry is the Cauchy-Riemann equations for the
the complex scalar $Z=X^6+iX^7$.
That is Z must be a holomorphic function of the complex worldvolume
coordinate $z=x^4+ix^5$. 
The calibration form for this intersection is the K\"{a}hler form. The
activated scalars in this case are $X^2$ to $X^7$.
\newpage
\bea
g_{2345}=g_{2367}=1\quad\quad\quad\quad\qquad \nu=1/4\quad {} \eea
\bea {X^2}' =-H^*\tb{3}{4}{5} -H^*\tb{3}{6}{7} &,& {X^3}' =
H^*\tb{4}{5}{2} +H^*\tb{6}{7}{2} \nonumber\\ {X^4}' =
-H^*\tb{5}{2}{3} &,& {X^5}' = H^*\tb{2}{3}{4}
\nonumber\\ {X^6}' = -H^*\tb{7}{2}{3} &,& {X^7}' =
H^*\tb{2}{3}{6}  \nonumber\\ \tb{2}{4}{6}= \tb{2}{5}{7} &,& 
\tb{2}{5}{6}=-\tb{2}{4}{7}  \nonumber\\ \tb{3}{4}{6}=
\tb{3}{5}{7} &,& \tb{3}{5}{6}=-\tb{3}{4}{7} \nonumber\\
\tb{4}{5}{6}=\tb{4}{5}{7}&=&\tb{4}{6}{7}=\tb{5}{6}{7}=0.
\nonumber \eea
In order to satisfy the constraint we need the $X^i$'s to satisfy the
following equations:
$$ \mbox{Choose } m\in\{2,3\},\quad i,j,k,l\in\{4,5,6,7\}\, ,\ \mbox{then}\nonumber\\
$$\bea \epsilon_{ijk}\fbr{i}{m}{m}{j}{k}&=&0,\quad(\mbox{no sum over }
m)\nonumber\\ \epsilon_{ijkl}\fbr{i}{j}{m}{k}{l}&=&0.\label{jac1} \eea

In the string theory case there were no additional equations, as
apart from the Nahm like equations and algebraic conditions on the
brackets all that was needed to solve the constraint was the Jacobi
identity, $\epsilon_{ijk}[\Phi^i,[\Phi^j,\Phi^k]]=0$. If $X^m$ anti-commutes with $X^i,X^j,X^k$ then
the first equation of (\ref{jac1}) reduces to the Jacobi identity. Similarly if $X^m$
anti-commutes with $X^i,X^j,X^k,X^l$ the second equation reduces to
\beq
 \epsilon_{ijkl}X^iX^jX^kX^l=0.
\eeq
For all the following configurations the additional algebraic
conditions take the same form as (\ref{jac1}). Note,  that although equations of this form are not
satisfied for general matrices in \mnc, it is possible that if the
$X^i$ are restricted to a particular algebra then these equations
could become identities. This suggests it is necessary to perhaps
restrict the algebra of fields on the membrane. We will return to
this idea later.

One can also have the following configurations as described in
\cite{berman1} with more and more complicated sets of
equations: three five-branes intersecting on a three-brane corresponding to an SU(3) K\"{a}hler
calibration of a two-surface embedded in six dimensions; three five-branes intersecting over a string 
which corresponds to an SU(3) K\"{a}hler calibration of a four-surface
in six dimensions; and three five-branes and an anti-five-brane
intersecting over a membrane which corresponds to the SU(3) special
Lagrangian calibration of a three-surface embedded in six dimensions.
These are all of the configurations preserving $1/8$ of the membrane
supersymmetry. There exist additional calibrations preserving less
supersymmetry with more five-branes, which can be treated in the same manner. 

\subsection{Solutions}

One can solve the cases of intersecting five-branes by using multiple
copies of the Basu-Harvey solution. The first multi-five-brane case is solved by setting
\beq 
X^i(s)=\frac{i\sqrt{2\pi}}{M_{11}^{3/2}}\frac{1}{\sqrt{s}}H^i, 
\eeq
where the $H^i$ are given by the block-diagonal $2N\times 2N$ matrices
\bea
H^2 &=& \mbox{diag}\, (G^1,G^1)\nonumber\\ H^3 &=& \mbox{diag}\,
(G^2,G^2)\nonumber\\ H^4 &=& \mbox{diag}\, (G^3,0)\nonumber\\ H^5 &=&
\mbox{diag}\, (G^4,0)\nonumber\\ H^6 &=& \mbox{diag}\,
(0,G^3)\nonumber\\ H^7 &=& \mbox{diag}\, (0,G^4)\nonumber\\ H^* &=&
\mbox{diag}\, (G^5,G^5)\, , 
\eea
which are such that
\beq 
H^i+\frac{1}{2(n+2)}g_{ijkl}\frac{1}{4!}[H^*,H^j,H^k,H^l]=0.  
\eeq
This makes sure that the conditions following
from the generalised Basu-Harvey equation vanish. The remaining
conditions, following from the
supersymmetry transformation, are satisfied trivially as all terms in the three
brackets involved vanish for this solution. The first additional
algebraic equation of (\ref{jac1}) is 
satisfied for this solution and the second additional algebraic
equation is trivially satisfied as there are no non-zero products of five different $X^i$'s.

The more complicated cases follow easily for example the third
configuration is also given by
the block-diagonal $3N\times 3N$ matrices
\bea H^2 &=& \mbox{diag}\, (G^1,G^1,G^1)\nonumber\\ H^3 &=&
\mbox{diag}\, (G^2,G^2,G^2)\nonumber\\ H^4 &=& \mbox{diag}\,
(G^3,0,0)\nonumber\\ H^5 &=& \mbox{diag}\, (G^4,0,0)\nonumber\\ H^6
&=& \mbox{diag}\, (0,G^3,0)\nonumber\\ H^7 &=& \mbox{diag}\,
(0,G^4,0)\nonumber\\ H^8 &=& \mbox{diag}\, (0,0,G^3)\nonumber\\ H^9
&=& \mbox{diag}\, (0,0,G^4)\nonumber\\ H^* &=& \mbox{diag}\,
(G^5,G^5,G^5)\, . \eea
The other configurations have similar block diagonal solutions.

There will exist many solutions containing off-diagonal terms. 
These will describe configurations when the branes are no longer
flat. It would be fascinating to find and analyse such solutions
since then one would be describing curved five-branes using
{\it{non-Abelian}} membranes.

\section{Fuzzy Spheres, membranes and $Q^{3 \over 2}$}

So far, the membrane fields are taken to lie in the algebra
of complex $N\times N$ matrices, \mnc. This choice is not unique as
discussed in \cite{berman2}. 
An alternative that we shall now consider is \anstp, the algebra that
reduces 
to the classical algebra of functions on the sphere in the large-$N$
limit.
It is significantly more technically involved than \mnc, since it is
not closed under multiplication. 
The lack of closure implies it is necessary to project back into the algebra after
multiplying. 
The projection then leads to the algebra to become
nonassociative, though obviously the
nonassociativity disappears in the large-$N$ limit as it must to
reproduce the classical algebra of functions.

To see how the projection works, decompose the full \mncs
fuzzy three-sphere basis into a basis of SO(4) Young tableau. The
projection is a restriction to allowing only completely symmetric
diagrams, that is those with only one row. This is described in detail
in \cite{berman2,coplandthesis}.

The essential point is that the algebra ${\cal A}_n(S^3)$ has
a tantalising property. n refers to representation of SO(4) given by
$({{n+1} \over 4},{{n-1} \over 4})$ which is a representation of the
fuzzy sphere algebra.
The number of degrees of freedom is no longer given by $N^2$ as it is for
\mncs, but it is now given by $D=(n+1)(n+2)(2n+3)/6$ \cite{berman2}. 
Thus for large $N$ where $n \sim \sqrt{N}$ we have that
\beq
D\sim N^\frac{3}{2},
\eeq
exactly as expected for $N$ coincident membranes in the large $N$
limit. 

We can see intuitively why a fuzzy three sphere should
have such a scaling. Recall that the dependence of the radius
of the fuzzy sphere on the
number of membranes is given by:
\beq
R \sim \sqrt{Q_2} \, .
\eeq
A noncommutative space
is equipped with a natural ultraviolet cut off given by the
noncommutative scale. There is also an infrared
cutoff given by the sphere size. Simply summing over all the spherical
harmonics, ie. the modes on the sphere, with the cut-offs prescribed as
above yields the number of degrees of freedom. In
units where the noncommutativity scale is one (and the radius is large)
this sum yields $R^3$. Therefore, the number of modes on a fuzzy three sphere whose
radius scales as $\sqrt{Q_2}$ is:
\beq 
D=Q_2^{3 \over 2} \, .
\eeq
This means that one can interpret the $Q^{3/2}$
degrees of freedom corresponding to the non-Abelian membrane theory as coming
from modes on the fuzzy sphere.

The appearance of a nonassociative algebra on five-branes 
in the presence of background flux is described in
\cite{Hofman:2001zt}.
Here there is flux due to the ending of the membrane on the five-brane and so it would be 
natural for nonassociativity to appear as the membranes expand out to
form a five-brane. Later we will see nonassociativity arising from
another perspective when trying to supersymmetrise the Basu-Harvey equation. 

Since the anti-symmetric bracket
vanishes when we project onto symmetric representations, one may ask how the
Basu-Harvey equation may still hold?
One possible presciption is that the projection does not act inside
the bracket, which instead is thought of as an operator $[G_5,\cdot,\cdot, \cdot]:({\cal
  A}_n(S^3))^3\rightarrow{\cal A}_n(S^3)$. To understand this we should
remember that the Basu-Harvey equation 
(or rather the fuzzy sphere equation (\ref{FS3}) on which it is based)
is a quantised version of a higher Poisson bracket equation. 
If we were to fix one of the
coordinates in a Poisson Bracket, the derivatives acting on that coordinate would give 
zero and the bracket would not hold. 
However, if we evaluate the derivatives and then fix 
the coordinate, the bracket equation still holds. 
Here the anti-symmetric bracket essentially encodes the derivatives,
which is why the projection is not made inside the bracket.

\section{A Nonassociative Membrane theory?}

In the previous sections we have seen the appearance of
nonassociativity resulting from the algebra of the fuzzy three
sphere. Also there has emerged a sextic coupling of the proposed non-Abelian
membrane theory and a proposed supersymmetric variation of the
membrane theory. What was missing was a proper supersymmetric action
including Fermions containing the
sextic coupling described above and an appropriate supersymmetetry
variation. Supersymmetric theories have been studied for many years
and theories with the sort of coupling in (\ref{m2act}) were known not
to be supersymmetrisable. The membrane must be a
supersymmetric theory and so we must resolve the puzzle of how to make this
theory supersymmetric.

Bagger and Lambert \cite{baggerlambert} had the novel idea to
consider the membrane fields to be part of some nonassociative
algebra. As we will see below, the
presence of the nonassociativity allows theory to be made
supersymmetric. This is remarkable since this now opens up the
possibility of a whole new class of supersymmetric theories that are
only supersymmetric if the fields take values in some nonassociative
algebra.

The goal will be to have a supersymmetric theory with eight scalars for which the
condition for half the supersymmetry transformations to vanish will 
be the Basu-Harvey equation. That is, the BPS
equation of theory, as defined by configurations that preserve half
supersymmetry, will be the Basu-Harvey equation.

We will consider the eight scalars labelled by $X^I$, $I=1..8$ with Fermions, $\Psi$. The
supersymmetry transformations will be:
\bea
\delta X^I &=& i {\bar{\epsilon}} \Gamma^I \Psi \\
\delta \Psi &=& \pl_\mu X^I \Gamma^\mu \Gamma^I \epsilon + i \kappa
[X^I,X^J,X^K] \Gamma^{IJK} \epsilon 
\eea
Then the vanishing of the supersymmetry transformation on Fermions obeying:
\beq
\P_{M2} \Psi =\Psi \quad {\rm{and}} \quad P_{M5}\Psi=\Psi
\eeq
implies the Basu Harvey equation.

Now if $X^I$ took values in a Lie Algebra then the triple bracket
would be equivalent to the Jacobi identity and so would vanish. We
thus take $X^I$ to be values in a nonassoiciative algebra where the
Jacobi identity is not obeyed. Given the product on the algebra
$( . )$, then the associator is defined as:
\beq
<X^I,X^J,X^K> = (X^I . X^J) . X^K - X^I . (X^k . X^K) 
\eeq
The triple bracket is then defined as:
\beq
[X^I,X^J,X^K]= {1 \over{12}} < X^{[I}, X^J,X^{K]} >
\eeq

The closure of this algebra is interesting since it only closes up to
translations and a conjectured local symmetry transformation. This is in analogy
with the D2 brane where the supersymmetry algebra only
closes up to translations and gauge transformations.

The new {\it{gauge}} symmetry on the membrane, required to close its
supersymmetry algebra, is given by:
\beq
\delta X^I = 6 i \kappa v_{JK} [X^I,X^J,X^K]
\eeq
and
\bea
\delta \Psi &=& {{3 i \kappa} \over {16}} v_{KL}\Gamma^{KL}
\Gamma^{IJ} [X^I,X^J,\Psi] \nonumber \\
&-& {{9i\kappa} \over 8} v_{IJ} [X^I,X^J,\Psi] -6i \kappa v_{IK}
\Gamma^{JK} [X^I,X^J,\Psi] \nonumber \\
&+& {{3 i \kappa} \over {192}} v_{JKLMN}\Gamma^J \Gamma^I
\Gamma^{KLMN} \Gamma^P [X^I,X^P,\Psi] \, ,
\eea
where
\beq
v_{IJ} = i {\bar{\epsilon}}_1 \Gamma_{IJ} \epsilon_2 \, , \qquad
v_{JKLMN}= -i {\bar{\epsilon}}_1 \Gamma_{J} \Gamma_{KLMN} \epsilon_2
\, .
\eeq
The meaning of these transformations is still uncertain. Obviously one would
require some sort of connection to make these transformations local but there
cannot be any more local degrees of freedom so the gauge field would need to
be be purely topological. This is still an open
question{\footnote{Current as yet unpublished work by Niel Lambert may
    throw further light on this question.}}.

The action which possess this supersymmetry is most easily
expressed in superspace formalism. In what follows we construct a
Lagrangian that is invariant under four of the sixteen
supersymmetries. Only the $SU(4) \times U(1)$ subgroup of the $SO(8)$ R-symmetry
will be manifest. This is the best that one can do (as yet), the full $SO(8)$
invariant version is still not known. It is conjectured that this is
because the above unknown gauge symmetry is not understood and
to get the full $SO(8)$ invariant theory will require the gauge
field. This is analogous to ${\cal{N}}=4$ Yang-Mills where in the purely
scalar-spinor sector one can only see a $SU(3) \times U(1)$ subgroup of the
SO(6) R-symmetry.
The fields transform under the $SU(4) \times U(1)$ symmetry as follows:

\begin{eqnarray*}
X^I && \rightarrow Z^A \oplus Z_{\bar{A}} \in {\bf{4}} (1) \oplus
{\bar{ \bf{4} } } (-1)  \\
\Psi && \rightarrow \psi^A \oplus \psi_{\bar{A}} \in {\bf{4}}(-1) \oplus
{\bar{\bf{4}}}(1)  \\
\epsilon && \rightarrow \epsilon \oplus \epsilon^* \oplus \epsilon^{AB}
\in {\bf{1}}(-2) \oplus {\bar{\bf{1}}}(2)\oplus {\bf{6}} (0)   \, .
\end{eqnarray*}  

Restricting to supersymmetries generated by $\epsilon$ ie. the N=2
algebra defined by $\epsilon^{AB}=0$, gives:
\beq
\delta Z^A=i {\bar{\epsilon}} \psi^A \, \, , \qquad \delta \psi^A=
\gamma^\mu \pl_{\mu} Z^A \epsilon +i \kappa_1 \epsilon^{ABCD}
[Z_{\bar{B}},Z_{\bar{C}},Z_{\bar{D}}] \epsilon^* +3 i \kappa_3 [Z^A,Z^B,Z_{\bar{B}}]\epsilon
\eeq
This corresponds to imposing: 
\beq
\Gamma_{5678}\epsilon= \Gamma_{56910}\epsilon =- \epsilon  \, .
\eeq
We can then interpret these projections as that corresponding to the SU(3)
Kahler calibrated geometry described in \cite{berman1}.

Thus this set up seems to be related to a particular calibrated
five-brane configuration. How one may relate this to other five-brane calibrations is still not known.
One must also note that the above supersymmetries do not close and so 
will require an additional constraint. This is most easily written in 
the superspace formalism which we now come to. Using the chiral superfield,
\beq
{\cal{Z}}^A=Z^A(y) +{\bar{\theta}}^*\psi^A(y) +{\bar{\theta}}^* \theta F^A(y)
\eeq
where $y^\mu=x^\mu + i \bar{\theta} \gamma^\mu \theta$, 
the action may be expressed as:
\beq
S=\int D^4\theta tr ({\cal{Z}}^A {\cal{Z}}_{\bar{A}} ) + \int D^2\theta
  W({\cal{Z}}) + \int D^2 {\theta}^* {\bar{W}}({\cal{Z}}_{\bar{A}})
\eeq
where the superpotential $W(Z)$ is given by:
\beq
W({\cal{Z}})= - {\kappa \over 8} \epsilon_{ABCD} Tr({\cal{Z}}^A,[{\cal{Z}}^B,{\cal{Z}}^C,{\cal{Z}}^D])           \, \, .
\eeq
To make this work we need a nonassociative algebra since if
${\cal{Z}}^A$ were Lie algebra valued then this superpotential would
vanish. This is described in the next section.

We must also supplement this with the superspace constraint
\beq
[{\cal{Z}}^A,{\cal{Z}}^B,{\cal{Z}}_{\bar{B}}]=0 \, \, ,
\eeq
which ensures the closure of the supersymmetry algebra.

\subsection{The nonassociative algebra}

We assume that the algebra is equipped with a bilinear trace form:
\beq
Tr: {\cal{A}}  \times {\cal{A}} \rightarrow C
\eeq
which obeys:
\beq
Tr(A, B)=Tr(B,A) \qquad  Tr(A . B, C) = Tr (A, B . C)
\eeq
There is also a complex involution of the algebra,
\beq
\#: {\cal{A}} \rightarrow {\cal{A}}
\eeq
such that $\#^2=1$ and
\beq
Tr(A,A^{\#} )\geq 0
\eeq
with equality only when $A=0$. Complex conjugation in the algebra is
then given by $\#$.

For a Lie algebra such as SU(N) there is an associative
anti-symmetric product  $i[ , ]: {\cal{A}} \times {\cal{A}} \rightarrow
 {\cal{A}} $ which preserves Hermiticity. What is required
 here is a triple  map that preserves the Hermitian condition. That is
 a trilinear map $[,,]: {\cal{A}} \times {\cal{A}} \times{\cal{A}}
 \rightarrow {\cal{A}}$ that obeys:
\beq 
( i [ X^I,X^J,X^K])^{\#} = i [X^I,X^J,X^K]  \, \, .
\eeq

We will now produce an explicit example of a nonassociative
algebra. Take $N \times N$ matrices but with a product defined as follows:
\beq
A . B = Q AB Q
\eeq
Where Q is a constant invertible matrix and the right handside is the
usual matrix product. The trace is defined by:
\beq
Tr(A,B)=tr(Q^{-1}AQ^{-1}B)
\eeq
where tr is the usual matrix trace. The generalised conjugation is
given by:
\beq
A^{\#}= QA^{\dagger} (Q)^{-1}{}^{\dagger}
\eeq
The associator of this algebra is:
\beq
<A,B,B,C>= Q^2 AB QCQ-QAQBCQ^2 \, \, .
\eeq
To eventually yield the Basu-Harvey equation we take:
\beq
Q={{1 + i G_5} \over \sqrt{2}}  \, .
\eeq
It is curious to see how $G_5$ now plays a key role in the nonassociative
algebra.
Using the above we may write the superspace action as:
\beq
S={1 \over 2} \int d^4 \theta  Tr({\cal{Z}}^A {\cal{Z}}_{\bar{A}} ) +
 - {\kappa \over 8} \int d^2 \theta \epsilon_{ABCD} Tr(G_5
 {\cal{Z}}^A,{\cal{Z}}^B{\cal{Z}}^C{\cal{Z}}^D]) + {\rm{h.c.}}        
\eeq
and then the BPS equation becomes:
\beq
\pl_\sigma Z_{\bar{A}} + {{\kappa} \over {4!}}
\epsilon_{ABCD}[G_5,Z^B,Z^C,Z^D]=0
\eeq
which we recognise as the Basu-Harvey equation with the parameter
$\kappa$ identified appropriately.

The suggestive punch line, however, is the counting of degrees of
freedom. Although the above representation uses $N \times N$
matrices for which we would d expect $N^2$ degrees of freedom, the algebra
allows us to truncate the matrices to only $N^{3/2}$ degrees of freedom. To see
this, consider $N \times N$ matrices with $N=n^2$ of the form:
\beq
 X= \sum_k^{n-1} X_k \otimes \Omega^k \, \, ,
\eeq
where $\Omega$ is an $n\times n$ matrix, such that $\Omega^n=1$. Thus X
has $n^3=N^{3/2}$ degrees of freedom.
Using the multiplication rule defined above with $Q=\Omega^q \otimes 1$
for some integer q
we obtain a nonassociative algebra with the appropriate degrees of
freedom. Taking $q=n/4$ and $G=\Omega^{n \over 2} \otimes 1$ we
reproduce the Basu-Harvey equation.

This nonassociativity is of course different fundamentally to that
discussed previously in the context of fuzzy three spheres. There the
nonassociativity and $N^{3/2}$ arose naturally but the system was not
supersymmetric. Here the supersymmetry is manifest and naturally
requires the nonassociativity but the $N^{3/2}$ is a
possibility that is imposed ad hoc. 

There are numerous open questions concerning this formalism. Perhaps
the most important is the
aforementioned local symmetry required by the closure of the
supersymmetry algebra. Understanding the role of
this symmetry is crucial in determing whether Bagger and Lambert's
conjectured membrane action has a chance of being related to the
non-Abelian membrane.

One crucial observation is the appearance of ${1\over \sqrt{N}}$ in
the Basu-Harvey equation and by extension also to $\kappa$ in the Bagger
Lambert formulation of the nonassociative membrane. 
This indicates that theory we are discussing is likely to be a
theory in the large N expansion. Of course in the
previous discussion of validity of the Basu Harvey equation we saw
how this was also the realm of validity of the fuzzy funnel solution. 
Given that the membrane is a theory whose coupling is outside the
quantum perturbative regeme this explains why one can have some sort
of simple semiclassical formalism. Carrying out a ${1 \over N}$ expansion
provides a possible perturbative parameter for theory. More work
on the ${1 \over N} $ expansion for the D2 brane seems to provide a hope to
understand to how the Basu-Harvey equation from the string theory point of view.

\section{Other ideas}

In the previous sections we
have described the basic problems of constructing theories with the
appropriate degrees of freedom that satisfy the properties which we 
determined via various circuititous means. In this final section we
will describe some other ideas and issues without entering into the details. 

One might wish to take the view that the five-brane is simply
the strong coupling limit of D4 brane and so should have a description
in terms of a 5 dimensional Yang-Mills theory or even, after further
compactifiaction, a four dimensional gauge theory. From examining the
supergravity description one can determine when the cross over from
the D4 to the M-theory five-brane should occur. This scale when
interpreted from the Yang-Mills point of view may be associated to 
fractional instantons\footnote{This idea was expressed
  to me by Tong and attributed to Arkani-Hamed.}. 
Thus the extra degrees of freedom in the five-brane theory may have an
interpretation in terms of fractional instantons that have become
in some sense {\it{light}}. A similar idea to this is the so called
deconstruction approach where one has the (0,2) theory arise out
of a limit of a lower dimensional quiver gauge theory \cite{deconstruct}.
Both these ideas provide interesting connections to gauge theories
but don't reveal much more of the strict M-theoretic description.

One missing piece of information, that would throw light onto the 
self-dual string, would be a supergravity
description of the membrane ending on the five-brane. There have been many 
attempts at this problem. Yet still a solution is far from being constructed. The reason for this is
as follows. One would be tempted to impose the supersymmetry projector
for the five-brane and membrane simultaneously and with a $SO(1,1)
\times SO(4) \times SO(4)$ metric ansatz, look for brane
solutions. This does not lead to the membrane ending on a five-brane
but actually to solutions of smeared insecting five-branes all be it
carrying membrane charge \cite{sugram2m5}. The reason for this is that
imposing a projector in supergravity essentially imposes an isometry
in those directions; this prevents a priori the solution from having
the properties we are looking for. There cannot be any metric dependence on
either brane world volume. This is in clear contradiction to what
we would expect from the self-dual string. One possible solution to this
is to consider deforming the known supersymmetry brane projectors so
as to remove the presence of some Killing directions. This is
essentially the type of solution constructing technique 
developed in \cite{chetham}\footnote{The author is
  greatful to Chethan Gowdigere for discussion on this question.}.
There has, however, been recent progress in finding brane solutions
in M-theory $AdS$ spaces, \cite{lunin}. In this work, solutions 
are found with $SO(2,2) \times SO(4) \times SO(4)$ symmetry. This is precisely the
symmetries one would expect for the supergravity dual of the 
decoupled self-dual string; the SO(2,2) being the isometry group of
$AdS_3$ and the $SO(4) \times SO(4)$ being the R-symmetry group of the
string. Following this there has been important work in determining
supergravity solutions of strings ending on branes and indeed the
M-theory analogue of the membrane ending on the five-brane
\cite{lunin2}. There have also been recent studies of the five-brane
wilson surface operators using AdS/CFT \cite{chen1} aswell as further
studies of the self dual string soliton in $AdS_4$ \cite{chen2}.

One approach to the five-brane theory is simply to take it as a theory
of a tensor multiplet and see how to construct a non-Abelian version. 
There are well known obstructions to this program. A clear no go
theorem is expressed in \cite{sevrin} which details the
inconsistencies of making a two form connection non-Abelian. However
one may take the following view. A two form potential may be used to
form a connection on loop space. Explicitly, for coordinates on the space
of loops $X^{\mu}(\sigma)$, the loop space covariant derivative is:
\beq
{\cal{D}}(\sigma)_\mu = {\delta \over {\delta X^{\mu}(\sigma)}} +
B_{\mu \nu}(X^{\mu}(\sigma)) \pl_\sigma X^\nu(\sigma)  \, .
\eeq
The commutator of covariant derivatives yields a field strength:
\beq
[{\cal{D}}(\sigma)_\mu,{\cal{D}}(\sigma')_\nu]= H_{\mu \nu \rho} \pl
X^{\mu}(\sigma) \delta(\sigma-\sigma')
\eeq
where $H=dB$, is the field strength of B, the Abelian two form connection. 
So one idea has been to generalise, not the two form connection,
but the loop space connection. The no go theorem then states that
this loop space connection cannot be expressed as an ultra local pull
back of some target space field. The appearance of the deformation of
the loop space by the presence of the background $C$ field on the brane
indicated that this may well be a way to view the five-brane. 
Ideas of this sort have been explored in \cite{gustavson,gustavson2} though 
as yet there is no clear theory of the five-brane.

This article on M-theory is missing a key section on the Matrix models of
M-theory \cite{matrixmodels}. This is because in matrix theory the
hardest things to explain seem to be the M-theory branes themselves,
never mind their interactions. There has been some clear progress \cite{MaldaMatrixM5}
on this issue, but gaining an insight into the $N^3$ or $N^{3/2}$ degrees of freedom
or the membrane five-brane interaction seems to be out of the realm of
matrix theory at the moment.

And so where are we at?
This review has described the evidence for new light
degrees of freedom in M-theory.
There are hints that coincident membranes may be described (perhaps in the large N limit) by
some nonassociative field theory. However, many mysteries remain to be
solved concerning aspects of M-theory brane interactions.
There are clues to new and exciting theories with novel
properties; to solve M-theory
questions we must move beyond examining the usual field theory suspects.

\section*{Acknowledgements}

The author is funded by EPSRC advanced fellowship grant GR/R75373/02 
and is grateful to DAMTP, Clare Hall College Cambridge and Rome, Tor
Vergata for hospitality during completion of this work. The author
also acknowledges the support of the Marie Curie research training
network MRTN-CT-2004-512194.
The author is very happy to acknowledge numerous conversations over
the years on M-theory and related topics with the following, some of
whom were collaborators and some just happy to share their
ideas (it goes without saying that all misunderstandings are due to the
author): Anirban Basu,  Eric Bergshoeff, Martin
Cederwall, Neil Copland, Nick Dorey, Jan de Boer,
Robert Dijkgraaf, Michael Duff, Shmuel Elitizur, Gary Gibbons, Michael Green, Jeff
Harvey,  Mans Henningson, Chris Hull, Finn Larsen, Neil Lambert, Ki
Myeoung Lee, Lubos Motl, Bengt Nielson, Neils Obers, Boris Pioline,
Malcolm Perry, Eliezer Rabinovici, Sanjaye Ramgolaam, Adam
Ritz, Jan Pieter Van der Schaar, Sav Sethi, Andrew Strominger,  
Per Sundell, David Tong, Paul Townsend, Erik Verlinde and Peter West.

\end{document}